# Pelletization Pressure Engineering of the Superconducting Properties of Fluorine-Doped PrFeAsO

Jayant Kumar[1], Konrad Kwatek[2], Tatiana Zajarniuk[3], Taras Palasyuk[2], Yaroslaw Konopelnyk[3], Tomasz Cetner[1], Cezariusz Jastrzębski[2], Shiv J. Singh[1]*

[1]*Institute of High Pressure Physics (IHPP), Polish Academy of Sciences, Sokołowska 29/37, 01-142 Warsaw, Poland*

[2]*Faculty of Physics, Warsaw University of Technology, Koszykowa 75, 00-662, Warsaw, Poland*

[3]*Institute of Physics, Polish Academy of Sciences, Aleja Lotnikow 32/46, 02-668 Warsaw, Poland*

***Corresponding author:**

Email: sjs@unipress.waw.pl

https://orcid.org/0000-0001-5769-1787

# Abstract

We investigate the influence of the pelletization pressure on the structural, microstructural, and superconducting properties of fluorine-doped $PrFeAsO_{0.7}F_{0.3}$ (Pr1111). Precursor powders are pressed into pellets at pressures ranging from 174 to 1131 MPa, after which the applied pressure is released and all samples undergo an identical two-step conventional synthesis process at ambient pressure (CSP-AP). All samples predominantly exhibit the tetragonal *P4/nmm* structure, and the lattice parameter *a* and unit-cell volume *V* decrease with increasing pelletization pressure up to an intermediate-pressure range (~900 MPa) before increasing at higher pressures. The actual fluorine-content derived from energy-dispersive X-ray spectroscopy (EDX) also increase with pelletization pressure, while Raman spectra show only modest variations in the characteristic vibrational modes associated with Fe and As. The superconducting response exhibits a non-monotonic dependence on the pelletization pressure. Among the investigated samples, the sample pelletized at 870 MPa exhibits the most favorable overall superconducting properties, with a superconducting onset temperature of 44.2 K, ~5 K higher than that of the lowest-pressure sample. It also displays a narrower superconducting transition, the highest relative sample density, a favorable effective magnetic critical current density at high fields, and a relatively large effective thermally activated flux-flow activation energy. At higher pelletization pressures, the superconducting response deteriorates, coinciding with increases in the lattice parameters, increased secondary-phase contributions, and reduced relative density. These findings establish the pelletization pressure as a relevant processing parameter for modifying the structural, microstructural and superconducting properties of Pr1111 during CSP-AP.

# I. Introduction

The discovery of superconductivity at 26 K in fluorine-doped LaFeAsO [1] established iron-based superconductors (IBS) as an important class of high-temperature superconducting materials and stimulated extensive investigations of the interplay among crystal structure, magnetism, electronic structure, and superconductivity [2, 3]. Among the different families of IBS, rare-earth iron oxypnictides, *RE*FeAsO (1111; *RE* = rare-earth element; *RE*1111), exhibit some of the highest superconducting transition temperatures ($T_c$), exceeding 50 K in optimally electron-doped compositions [4]. In these compounds, fluorine substitution at the oxygen site introduces electron carriers into the superconducting FeAs layers and suppresses the antiferromagnetic state of the parent compound, leading to the emergence of superconductivity. Their superconducting properties are also closely associated with structural parameters such as the Fe–As bond length, pnictogen height, and $FeAs_4$ tetrahedral geometry [5, 3]. Consequently, controlling both the crystal structure and microstructure provides important opportunities for optimizing the superconducting properties of 1111-type materials.

PrFeAs(O,F) is a particularly interesting member of the 1111 family because of its relatively high superconducting transition temperature ($T_c$~50 K) and broad superconducting composition range [6, 7, 8]. Recent investigations of $PrFeAsO_{1-x}F_x$ over a wide fluorine concentration range ($0 \leq x \leq 1$) have established its electronic phase diagram and identified the optimal doped region around $x$ ~0.30–0.35, where $T_c$ reaches ~45 K [6]. $PrFeAsO_{0.7}F_{0.3}$ therefore provides a suitable composition for investigating the influence of processing conditions on superconductivity while keeping the nominal chemical composition fixed. In addition to the intrinsic electronic and structural properties, the performance of polycrystalline Pr1111 is strongly influenced by extrinsic factors, including phase purity, sample density, grain connectivity, defects, and vortex pinning. These factors are particularly important for the critical current density ($J_c$), which depends on both intragrain flux pinning and the efficiency of intergrain current transport [4, 9].

Pressure has been widely used to investigate and modify the structural, electronic, and superconducting properties of IBS. Externally applied pressure can alter interatomic distances, lattice parameters, and electronic and magnetic interactions, thereby producing composition-dependent changes in the superconducting transition temperature and other physical properties [10, 11, 12, 13] [14]. For example, pressure-dependent resistivity and magnetization measurements on $PrFeAsO_{0.6}F_{\gamma}$ have demonstrated that the pressure dependence of $T_c$ varies with fluorine concentration [13]. Such studies provide insights into the relationship between structural changes and superconductivity under externally applied pressure. In addition to measurements performed under applied pressure, pressure can also influence IBS when maintained during synthesis or crystal growth. When pressure is maintained during thermal processing, it can influence phase formation, chemical reactions, and the resulting phase assemblage and microstructure. High-pressure growth studies of iron-based oxypnictides have demonstrated that these processing conditions can modify the resulting phases and superconducting properties [15, 16, 17, 18]. In particular, high-pressure synthesis studies of $PrFeAsO_{1-x}F_x$ have demonstrated that the effects on fluorine incorporation, phase assemblage, microstructure, and superconducting properties depend on the nominal fluorine concentration [19]. Pressure-assisted densification techniques, such as spark-plasma sintering, have also been employed to modify the microstructure and superconducting properties of iron oxypnictides through the combined application of pressure and heat [20]. These approaches demonstrate the importance of pressure in superconducting-material research, although their effects depend on whether pressure is applied during property measurements, maintained during synthesis, or combined with thermal processing during densification.

A distinct approach is to apply pressure only during room-temperature pelletization of the precursor powders [21, 22]. During this process, applied pressure modifies the initial packing of the powders, particle contacts, and pore structure of the compact. The pressure is then completely released before heat treatment, and the subsequent thermal reaction takes place at ambient pressure. Nevertheless, differences in the initial compacted state may influence the subsequent reaction and microstructural evolution, potentially affecting the phase assemblage, compositional homogeneity, density, grain development, and superconducting properties of the final material. Thus, the influence of pelletization pressure may persist in the heat-treated material even though pressure is not maintained during phase formation. This approach differs fundamentally from high-pressure synthesis and crystal-growth methods, in which pressure is maintained during the thermal reaction or growth process and can directly influence the thermodynamic and kinetic conditions of phase formation. Although applied-pressure studies, high-pressure synthesis, and pressure-assisted densification [20, 23] have been investigated in iron-based superconductors, their results do not directly establish how room-temperature pelletization pressure, followed by complete pressure release and an identical ambient-pressure heat treatment, affects the final properties of fluorine-doped *RE*FeAsO or PrFeAsO. Differences in the initial compacted state may affect densification and intergranular connectivity, while subsequent reaction and microstructural evolution may be accompanied by changes in phase assemblage and compositional homogeneity. These factors may correlate with the superconducting transition, normal-state transport, vortex dynamics, and effective magnetic critical current density. A systematic investigation is therefore needed to establish the correlations among pelletization pressure, structural parameters, measured fluorine content, phase assemblage, sample density, and superconducting properties in fluorine-doped PrFeAsO.

In this work, we systematically investigate the effect of pelletization pressure on nominally fluorine-doped $PrFeAsO_{0.7}F_{0.3}$ polycrystalline samples. The precursor powder were compacted into disk-shaped pellets at the applied pressures ranging from 174 to 1131 MPa using a hydraulic press, after which the applied pressure was completely released. All pellets were subsequently subjected to the same two-step conventional synthesis process at ambient pressure (CSP-AP). The resulting samples were characterized by X-ray diffraction, scanning electron microscopy (SEM), energy-dispersive X-ray spectroscopy (EDX), Raman spectroscopy, electrical resistivity, magnetoresistance, and dc magnetization. These complementary techniques are used to examine the relationships between pelletization pressure and the structural, compositional, microstructural, transport, and magnetic properties, including vortex dynamics and effective magnetic critical current density. Among the investigated conditions, a pelletization pressure of 870 MPa yields the most favorable overall superconducting performance, with an onset transition temperature of 44.2 K, a relatively narrow transition, and favorable effective magnetic critical-current performance at high fields. At higher pelletization pressures, changes in structural parameters, secondary-phase contributions, and the final microstructure coincide with deterioration of the superconducting response. These results establish correlations between pelletization pressure and the final properties of $PrFeAsO_{0.7}F_{0.3}$ following CSP-AP.

## II. Experimental Details

Polycrystalline $PrFeAsO_{0.7}F_{0.3}$ samples were prepared using a conventional two-step solid-state reaction at ambient pressure, preceded by controlled pressure-assisted pelletization. High-purity starting materials, Pr (purity: 99.9%, powder), $PrF_3$ (purity: 99.9%, powder), $Fe_2O_3$ (purity: 99.99%, powder), Fe (purity: 99.99%, powder), and As (purity: 99.9999%, lumps), were weighed according to the nominal $PrFeAsO_{0.7}F_{0.3}$ composition, with a total batch mass of approximately 12 g. All powder-handling, weighing, mixing, and pelletization procedures were performed in an argon-filled high-purity glove box with oxygen and moisture concentrations maintained below 1 ppm to minimize contamination and oxidation. The starting materials were thoroughly mixed and ground in an agate mortar for ~45 min to promote compositional homogeneity. The resulting precursor mixture was divided into several portions of ~1.5 g, each of which was compacted into a pellet using an ~10.4 mm diameter steel die and a hydraulic press inside the high purity glove box. Further details of the pelletization procedure are provided in Section A of the Supplementary Information. Different pelletization pressures of 174, 348, 522, 696, 870, 1044, and 1131 MPa were applied to the individual pellets, resulting in seven samples designated S1–S7, respectively, as summarized in Figure 1 and Table ZT1 of the Supplementary Information. All pellets were prepared from the same precursor batch, ensuring identical nominal composition and initial chemical processing history. After compaction, the applied pressure was completely released, and no external pressure was maintained during either heat-treatment step. Consequently, the pressure investigated here refers exclusively to room-temperature pelletization rather than high-pressure synthesis. To limit fluorine loss and avoid chemical reactions with the quartz tube, the pellets were sealed inside tantalum (Ta) tubes, which were mechanically crimped at both ends. The Ta tubes were subsequently encapsulated in evacuated quartz tubes under a vacuum of $10^{-2}$–$10^{-3}$ mbar to minimize oxidation during heat treatment. The thermal treatment was performed under ambient pressure using a two-step solid-state reaction procedure established through previous experimental optimization of PrFeAs(O,F) phase formation [6, 24]. In the first step, the samples were slowly heated to 550 °C and held for 8 h, followed by furnace cooling to room temperature. This low-temperature pre-reaction step was intended to initiate the solid-state reaction while limiting the volatilization of fluorine-containing species. In the second step, the samples were reground and re-pelletized inside the glove box at the same pelletization pressure, resealed in Ta tubes, and again encapsulated in evacuated quartz tubes. The pellets were then annealed at 950 °C for 12 h, followed by controlled cooling to 110 °C over 18 h and subsequent furnace cooling to room temperature [6, 24]. Based on our experimental observations, additional or prolonged heat treatment can promote fluorine loss and may be accompanied by a decrease in the superconducting transition temperature [4, 25]. Therefore, the same two-step heat-treatment schedule was applied to all samples investigated in the present study. Accordingly, all structural, compositional, vibrational, transport, and magnetic characterizations reported in this work were performed on the samples after the prescribed two-step heat-treatment schedule. Pelletization pressure was therefore treated as the principal intentionally varied processing parameter, and its association with the structural, compositional, and superconducting properties of the final prepared samples was evaluated under otherwise identical thermal-processing conditions.

The phase composition and crystal structure of all prepared samples were examined by powder X-ray diffraction (XRD) using a PANalytical Empyrean diffractometer with Cu Kα radiation (λ = 1.5418 Å), operated at 40 kV and 35 mA. Diffraction patterns were collected in continuous scanning mode over the 10–90° $2\theta$ range with a nominal angular step of 0.013°, yielding around 6092 data points. The total acquisition time was ~63 min per pattern. Phase identification was performed using the PDF-4+ database (ICDD). Rietveld refinements of the powder X-ray diffraction data were subsequently carried out using the GSAS software package to determine the lattice parameters and estimate the weight fractions of the crystalline phases. Further details of the Rietveld refinement procedure are provided in Section B of the Supplementary Information. The microstructure and elemental composition of the samples were examined using a Zeiss Ultra Plus field-emission scanning electron microscope equipped with a Bruker Quantax 400 energy-dispersive X-ray spectroscopy (EDX) system with an Ultraspeed detector (30 $mm^2$ active area; nominal energy resolution 127 eV). These measurements were performed at an accelerating voltage of 15 kV. Measurements were also tested at 20 and 25 kV; however, 15 kV provided the most suitable and reproducible spectra for the present samples. Before scanning electron microscopy (SEM) examination, the pellets were polished using emery papers of progressively finer grades. The bulk density of each sample was determined from its measured mass and geometrical volume calculated from its dimensions as listed in Table ZT3 of the supplementary information. The sample dimensions were measured after heat treatment, and the bulk density was calculated as the ratio of mass to volume. The resulting densities are shown in Figure 1 and listed in Table ZT3 of the Supplementary Information. Because the dimensions were measured after heat treatment, the reported density represents the final bulk density of the heat-treated samples rather than the green density immediately after pelletization. The relative density of each sample was calculated by normalizing the experimentally measured bulk density to the theoretical density determined from the Rietveld-refined structural parameters of the corresponding sample. The resulting relative density therefore represents the final densification state after the prescribed two-step CSP-AP.

Raman scattering measurements were performed using a Horiba Jobin Yvon LabRAM ARAMIS spectrometer equipped with a 632.8 nm He–Ne laser as the excitation source, following the approach used in previous studies [26, 16, 17, 6]. The laser beam was focused onto the sample surface using a 100× objective lens with a numerical aperture of 0.95, producing a spot size of less than 3 μm. The scattered light was collected in a backscattering geometry, dispersed using a diffraction grating with 2400 grooves $mm^{-1}$, and detected with a thermoelectrically cooled charge-coupled device (CCD). Raman spectra were collected over the spectral range of 95–300 $cm^{-1}$. To minimize local laser heating, the incident laser power was maintained at ~140 μW, and each spectrum was accumulated for 300 s. Measurements were performed on polished surfaces of the as-synthesized samples S1 to S7. Because the samples were polycrystalline, spectra were collected from at least five different locations on each specimen to assess the reproducibility of the observed phonon modes. The Raman-active vibrational modes associated with the PrFeAs(O,F) phase were identified by fitting the experimental spectra using Lorentzian line shapes.

The magnetic properties were measured using a vibrating sample magnetometer (VSM) integrated into a Physical Property Measurement System (PPMS, Quantum Design). The temperature dependence of the magnetization was measured under both zero-field-cooled (ZFC) and field-cooled (FC) conditions in an applied magnetic field of 20 Oe over the temperature range of 5–60 K. These measurements were used to compare the superconducting magnetic response and transition behavior of the samples; no quantitative bulk superconducting volume fraction was determined from the ZFC/FC data. Magnetic hysteresis ($M$–$H$) loops were recorded at 5 K in magnetic fields up to 9 T with a magnetic-field sweep rate of 50 Oe per second. To account for the non-superconducting magnetic background, an $M$–$H$ loop was recorded at 60 K, above the superconducting transition temperatures of all samples, for each sample and subtracted from the corresponding 5 K loop. The background-corrected hysteresis loops were then used to estimate the effective magnetic critical current density, $J_c$, using the Bean critical-state model [27]. The same measurement and background-correction procedure was applied to all seven samples to provide a consistent basis for comparative evaluation. Because the samples are porous and polycrystalline, with significant porosity and possible weak intergrain connectivity, the Bean-model values are regarded as effective magnetic $J_c$ values rather than intrinsic transport critical current densities representative of fully dense polycrystalline material. Electrical transport measurements were performed using a conventional four-probe configuration on rectangular specimens. Silver paste and copper wires were used to establish electrical contacts. The temperature dependence of the electrical resistivity was measured over a temperature range extending from 5 K to 280-290 K under zero applied magnetic field using either a closed-cycle refrigerator or the PPMS system. Magnetotransport measurements were performed using the PPMS under applied magnetic fields of up to 9 T to investigate the field dependence of the superconducting transition and the associated vortex-transport properties. The vortex activation energy was determined from the temperature dependence of the resistivity in the thermally activated regime using the Arrhenius analysis [28].

# III. Results and Discussions

## (a) Structural Analysis

Figure 2(a) presents the powder X-ray diffraction (XRD) patterns of samples S1–S7 prepared using different initial pelletization pressures (Supplementary Table ZT1). The major diffraction reflections are indexed to the tetragonal PrFeAs(O,F) phase with space group *P4/nmm*, consistent with previously reported fluorine-doped PrFeAsO compounds [6, 7, 8, 24]. Weak additional reflections attributable to FeAs and PrOF secondary phases are also observed as similar to the previous reports [7, 6, 29]. The PrOF phase remains nearly unchanged over the applied pelletization pressure range from 174 to ~1044 MPa (S1–S6), whereas a slight increase is observed for the highest-pressure sample S7. The FeAs impurity phase exhibits a slight increase with increasing pelletization pressure, becoming more pronounced at higher pressure. The pressure-dependent evolution of the estimated secondary-phase fractions is summarized in the supplementary Figure Z1. These results show that the phase assemblage of the final heat-treated samples is associated with the applied pelletization pressure. From 174 to 1044 MPa, PrOF fraction remained nearly constant, while the FeAs fraction increased slightly. At further increase in pelletization pressure i.e. the sample S7, the increased contributions from secondary phases indicate a further change in the final phase assemblage.

To obtain quantitative estimates of the structural parameters, Rietveld refinements were performed for all samples. A representative refinement for sample S5 is shown in Fig. 2(b), while the corresponding refinements for the other samples are provided in the supplementary Figure Z2(a–f). The refined structural parameters, phase fractions, and refinement-quality indicators ($R_p$, $R_{wp}$, and $\chi^2$ ), are now provided in the Supplementary Table ZT2. The refined lattice parameters $a$ and $c$, together with the unit-cell volume $V$, are summarized in Figures 2(c–e). As shown in Figure 2(c), the lattice parameter $a$ show a slight overall decrease with increasing pelletization pressure and reaches a minimum for samples S5 and S6, prepared at ~870 and 1044 MPa, respectively. A further increase in the pelletization pressure results in a slight increase in lattice $a$ for the sample S7. In contrast, the lattice parameter $c$ exhibits only relatively small variations over the investigated pressure range (Figure 2(d)), suggesting that the observed structural variations are more pronounced in the *ab* plane than along the crystallographic $c$ direction. Consequently, the unit-cell volume $V$ decreases overall from S1 to S5, remains nearly unchanged between S5 and S6, and increases slightly for the sample S7 (Figure 2(e)). The magnitude of these changes may be comparable to the uncertainties in the refined structural parameters or may reflect small variations in the actual fluorine incorporation into the superconducting Pr1111 phase.

To examine the variation of the EDX-derived average fluorine content $x_{act}$ with the applied pelletization pressure, energy-dispersive X-ray spectroscopy (EDX) measurements were performed on polished surfaces of all samples. EDX analysis has previously been employed for compositional characterization of fluorine-doped 1111-type iron-based superconductors, including $PrFeAsO_{0.60}F_{0.12}$ and F-doped SmFeAsO systems [29, 30]. In the present study, EDX is used primarily to compare compositional trends among samples measured using the same instrument and under identical analytical conditions. Figure 2(f) presents the EDX-derived average fluorine content, $x_{act}$, as a function of pelletization pressure. For the low-pressure samples S1–S3, $x_{act}$ is ~0.13, substantially below the nominal fluorine concentration of 0.30. With increasing pelletization pressure, the $x_{act}$ increases progressively, reaching ~0.19 for the sample S6. For the highest-pressure sample S7, the EDX-derived fluorine content $x_{act}$ shows little further increase, indicating an apparent saturation of the measured fluorine content under the present processing conditions. Although identical measurement conditions provide a basis for comparative analysis, differences in porosity, surface condition, matrix composition, and phase assemblage may affect the accuracy of the EDX-derived $x_{act}$ values. Accordingly, $x_{act}$ is treated as a comparative compositional indicator rather than an absolute measure of fluorine occupancy in the superconducting Pr1111 phase. As shown in Supplementary Fig. Z1, the PrOF phase fraction remains approximately constant across samples S1–S6 but increases in S7. Since EDX provides an average composition of the analysed regions, its fluorine estimate may include contributions from both the PrFeAs(O,F) phase and fluorine-containing secondary phases such as PrOF. For samples S1–S6, the approximately constant PrOF phase fraction, together with the observed variation in the EDX-derived $x_{act}$ values, may be considered indicative of changes in the fluorine distribution within the samples. These variations are accompanied by decreases in lattice parameter $a$ and unit-cell volume $V$, suggesting a possible correlation between the EDX-derived fluorine content and the structural evolution of the Pr1111 phase.

To further examine the relationship between the EDX-derived actual fluorine content $x_{act}$ and structural evolution, the lattice parameter $a$ and unit-cell volume $V$ of our samples are plotted against $x_{act}$ and compared with previously reported data for F doped Pr1111 [6], as shown in Supplementary Figure Z3(a)-(b). For samples S1–S5, the observed reduction in $a$ and $V$ with increasing pelletization pressure is accompanied by an increase in the EDX-derived average fluorine content $x_{act}$ of the analysed regions. Fluorine substitution for oxygen can produce lattice contraction because of differences in ionic size and the resulting changes in the crystal structure [6, 7, 8]. For samples S1–S5, both $a$ and $V$ decrease with increasing $x_{act}$, following the general trend reported in Ref. [6]. This agreement is qualitatively consistent with a possible increase in fluorine incorporation into the superconducting Pr1111 lattice over this pressure range; however, the EDX measurements alone cannot establish this incorporation quantitatively. However, for S6 and S7, the lattice parameter and unit-cell volume increase with slight increase overall EDX-derived fluorine content $x_{act}$. This expansion is more pronounced in the sample S7, where the PrOF phase fraction also increases (Supplementary Figure Z1). The subsequent expansion of $a$ and $V$ at the highest pelletization pressure indicates that the structural evolution is no longer monotonic at the highest pressure. These observations suggest that, at higher pelletization pressures (> ~1000 MPa), an increasing fraction of the fluorine may reside in the PrOF secondary phase rather than being incorporated into the superconducting Pr1111 lattice. Overall, the correlations among $x_{act}$, lattice parameters, and phase composition suggest that pelletization pressure influences fluorine incorporation in the superconducting Pr1111 lattice and phase formation of Pr1111 during subsequent heat treatment. The trends suggest that fluorine incorporation into the Pr1111 lattice may be favored up to an intermediate pelletization pressure (~900 MPa), whereas further pressure increase is accompanied by enhanced secondary-phase PrOF formation. A similar result has been observed for high pressure synthesis F-doped Sm1111 by the high gas pressure and high temperature synthesis technique [16]. However, because EDX cannot distinguish fluorine in the superconducting phase from that in secondary phases, these results do not establish the absolute fluorine occupancy of the Pr1111 lattice or provide a quantitative fluorine mass balance. The observed trends should therefore be interpreted as evidence of coupled compositional and structural evolution rather than as a direct measurement of fluorine incorporation. In the supplementary figure Z5, the data presented in Figures 2(c)-(e) are shown without guide to the eye lines.

The bulk density provides further insight into the effect of the applied pelletization pressure on the samples S1 to S7. The sample density was calculated from the measured sample mass and geometrical volume determined from each rectangular shaped sample dimensions (Supplementary Table ZT3). The relative density is obtained by normalizing the experimental bulk density to the theoretical density calculated from the Rietveld-refined structural parameters of each sample (Supplementary Table ZT3). The resulting relative density is plotted as a function of pelletization pressure in Figure 1. The relative density increases rapidly with pelletization pressure, reaching a maximum for the sample S5 at 870 MPa, followed by a decrease at higher pressures. This non-monotonic trend indicates that increasing the initial pelletization pressure does not lead to continuous densification after the subsequent heat treatment, with the highest relative density obtained near the S5 processing condition. Further increase in the pelletization pressure does not result in additional densification and is instead accompanied by a decrease in the final relative density. The decrease at the highest pressure occurs together with the increased presence of secondary phases (Supplementary Figure Z1) and changes in the EDX-derived average fluorine content $x_{act}$. The correspondence between the maximum relative density for the sample S5 processing condition and the subsequent improvement in superconducting properties, as discussed below, observed for this sample suggests that the final density and associated intergrain connectivity may contribute to the superconducting response [31, 8].

## (b) Microstructural Analysis

All pellets were polished using emery papers of progressively finer grades inside a high-purity Ar-filled glovebox, without using any oil, to minimize contamination during sample polishing. Representative backscattered-electron (BSE) SEM micrographs of samples S1, S5, and S7 at different magnifications are presented in Figures 3(a–i), while images of the remaining samples are provided in Supplementary Figures Z6(a–l). In the BSE micrographs, the light-gray matrix is attributed to the PrFeAs(O,F) target phase, whereas brighter regions are attributed to Pr-rich oxide/oxyfluoride phase PrOF($Pr_2O_3$). Dark regions may represent pores and/or FeAs phase. These assignments are consistent with the phase composition identified by XRD. Such microstructural features are commonly observed in polycrystalline 1111-type IBS synthesized by CSP-AP [4, 16, 19]. Sample S1 exhibits a relatively porous and heterogeneous microstructure, with large grains, numerous voids and distinct secondary-phase regions. The presence of secondary phases is consistent with the XRD identification of PrOF and FeAs (Figures 3(a–c)) and the observed porosity is consistent with its relatively low relative sample density, as shown in Figure 1. As the pelletization pressure increases, samples S2–S4 exhibit a comparatively more compact microstructure, with fewer apparent voids than S1, suggesting improved particle packing and interparticle contact. This observation agrees with the increase in relative sample density (Figure 1). Sample S5, prepared at an intermediate pelletization pressure of 870 MPa, exhibits the most compact microstructure among the investigated samples, although residual pores and secondary-phase regions remain visible (Figures 3(d–f)). At higher pelletization pressures, the microstructure becomes less uniform. In S6, pores are more apparent (Supplementary Figures Z6(j–l)), while S7 exhibits more frequent pores and secondary-phase regions (Figures 3(g–i)). These features are consistent with the decrease in relative density at higher pelletization pressures i.e., for S7 (Figures 3(g–i)) and the existing of pores and the secondary-phase regions are observed more frequently. The increased presence of secondary phases, including PrOF and FeAs, in S7 is consistent with the XRD results and the decrease in relative sample density. Overall, the microstructural evolution indicates that the beneficial effect of increasing pelletization pressure reaches an optimum under the investigated conditions at ~870 MPa. The comparatively high relative density and compact microstructure of S5 may favour macroscopic current transport by reducing the detrimental effects of porosity and potentially improving intergrain connectivity. Improved connectivity may help mitigate weak-link effects that commonly limit the current-carrying performance of polycrystalline iron-based superconductors [9]. Conversely, the increased porosity and presence of secondary phases observed at higher pressures may adversely affect current transport. Hence, the superconducting properties of these samples should be influenced by the combined effects of relative sample density, grain connectivity, and phase composition.

## (c) Raman Spectroscopy

Representative Raman spectra of the $PrFeAsO_{0.7}F_{0.3}$ samples pelletized at different initial pressures and subsequently heat-treated under identical ambient-pressure conditions are presented in Figure 4(a). Three dominant signals are observed in the investigated spectral range at ~164 $cm^{-1}$, 204 $cm^{-1}$ and 209 $cm^{-1}$ and are assigned to the Raman-active collective lattice vibrations (phonon modes) of $A_{1g}$, $A_{1g}$, and $B_{1g}$ symmetry, respectively. These modes predominantly involve the out-of-plane displacements of the Pr, As, and Fe atoms, and are denoted as the Pr($A_{1g}$), As($A_{1g}$), and Fe($B_{1g}$) modes, respectively. The mode assignments are also consistent with our previous investigations of PrFeAs(O,F) [6, 24, 32] as well as with reported Raman studies of other rare-earth 1111 IBS [33, 34, 35]. The phonon modes observed for all samples are indicated in Figure 4(a). Because the samples are polycrystalline, the relative intensities of the Raman modes can be affected by the random orientation of individual crystallites. Therefore, the analysis here focuses primarily on the phonon frequencies, which provide information about the local bonding environment and lattice dynamics. The pelletization-pressure dependence of the Raman shifts of the investigated Raman-active modes is summarized in Figure 4(b). The Fe $B_{1g}$ and As $A_{1g}$ modes show a slight overall increase in frequency with increasing pelletization pressure, particularly from the lowest-pressure samples toward the intermediate-pressure range. In contrast, the Pr $A_{1g}$ mode exhibits a more non-monotonic variation without a clear systematic trend. The magnitude of these frequency variations is relatively small compared with the corresponding phonon frequencies, indicating that the measured lattice vibrational response changes only modestly across the pelletization-pressure series. The observed Raman evolution is considered alongside the structural and compositional changes obtained from XRD and EDX. In particular, the slight increase in the As $A_{1g}$ and Fe $B_{1g}$ frequencies toward the intermediate-pressure range occurs together with a decrease in the lattice parameter $a$ and unit-cell volume $V$ and an increase in the observed EDX-derived average fluorine content $x_{act}$ as discussed above. Since fluorine substitution at the oxygen site can modify the local bonding environment and is associated with lattice contraction in PrFeAs(O,F), the concurrent evolution of the lattice parameters, the EDX-derived average fluorine content $x_{act}$, and Raman-active phonon frequencies provides complementary qualitative evidence of pelletization-pressure-dependent changes in the local structural and chemical environment of Pr1111. These trends seem consistent with the changes in fluorine incorporation into the PrFeAs(O,F) lattice, although the present measurements do not independently establish this incorporation. Therefore, the observed Raman shifts should be regarded as complementary qualitative evidence of modest changes in the vibrational response, rather than as a direct quantitative measure of fluorine occupancy in the Pr1111 lattice.

## (d) Transport Properties

The temperature dependence of the electrical resistivity, $\rho(T)$, was measured under zero magnetic field for the samples S1–S7 to investigate the evolution of normal-state transport and superconductivity with the applied pelletization pressure. The resistivity curves over the full temperature range are shown in Figure 5(a), while the low-temperature region around the superconducting transition is presented in Figures 5(b). All samples exhibit metallic-like behavior over most of the investigated temperature range, with a nearly linear temperature dependence of resistivity in the normal state i.e. above the superconducting transition. The absolute normal-state resistivity exhibits a pronounced but non-monotonic variation with the applied pelletization pressure, rather than a simple monotonic trend. This non-monotonic behavior indicates that the normal-state resistivity cannot be described by a simple dependence on the pelletization pressure. It may reflect the combined effects of phase composition, EDX-derived average fluorine content, porosity, and intergrain connectivity. A pronounced dependence of the superconducting transition on the pelletization pressure is observed. For the lowest-pressure sample S1, the superconducting transition occurs at ~39 K. With increasing pelletization pressure, $T_c$ increases from S1 to S5 and reaches a maximum for sample S5, for which $T_c$ is ~44–45 K, corresponding to an enhancement of about 5–6 K compared with the sample S1. This improvement coincides with the increase in the actual fluorine content $x_{act}$ experimentally determined by EDX, the contraction of the lattice parameters observed by XRD, and an increase in the sample density (Figure 1). These correlations suggest that differences introduced during the initial pelletization step are associated with changes in the actual fluorine content $x_{act}$, phase formation, and densification during the subsequent heat treatment, which may collectively contribute to the improved superconducting performance.

The superconducting transition deteriorates when the pelletization pressure is increased beyond the optimum range. For samples S6 and S7, pelletized at 1044 and 1131 MPa, respectively, $T_c$ decreases to ~43.4 K and ~37.7 K, respectively. This suppression occurs concurrently with the changes in the secondary-phase fractions and the reduction in the sample density at higher pelletization pressures. In particular, the increase in the estimated PrOF phase fraction at the highest pressure (~1131 MPa) (Supplementary Figure Z1), together with the apparent saturation of the EDX-derived average fluorine content (Figure 2(f)), suggests that further increases in the pelletization pressure do not result a corresponding increase in the actual fluorine incorporation into the superconducting PrFeAs(O,F) main phase. The increased PrOF fraction may be contribute to limiting the fluorine content retained in the main superconducting phase. The accompanying changes in phase inhomogeneity and the reduced sample density may adversely affect the continuity of intergrain current transport and the macroscopic superconducting response. The dependence of $T_c$ on the pelletization pressure therefore exhibits a distinctly non-monotonic behavior, with the most favorable superconducting response obtained for S5, pelletized at ~870 MPa. Importantly, this behavior cannot be explained simply by a monotonic change in the average crystal structure with the pelletization pressure. Raman measurements show only small pressure-dependent variations in the investigated phonon modes, with the Fe $B_{1g}$ and As $A_{1g}$ modes showing a slight overall increase toward the intermediate-pressure range, while the Pr $A_{1g}$ mode exhibits a more non-monotonic variation. Taken together with trends in relative sample density and secondary-phase content, these observations suggest that the improved superconducting properties at intermediate pelletization pressures are associated with the combined effects of fluorine incorporation, phase assemblage, and final microstructural quality, rather than with substantial changes in the measured phonon frequencies alone. At pressures above ~1000 MPa, the concurrent decrease in $T_c$, increase in secondary-phase contributions, and reduction in final sample density suggest that excessive pelletization pressure is unfavorable for achieving optimal superconducting properties in PrFeAs(O,F) after subsequent CSP-AP. The observed non-monotonic processing dependence identifies ~900 MPa as the most favorable pelletization pressure among the conditions investigated, where the sample exhibits the highest relative density and the best overall superconducting response. This processing-pressure dependence provides a link between the initial pelletization conditions and the superconducting properties of $PrFeAsO_{0.7}F_{0.3}$ after subsequent CSP-AP.

Figure 5(c)-(e) summarizes the pelletization-pressure dependence of the superconducting transition temperature $T_c^{onset}$, transition width $\Delta T$, and residual resistivity ratio ($RRR = \rho_{280K}/\rho_{50K}$) for samples S1–S7, respectively. The transition width was calculated as $\Delta T = T_c^{onset} - T_c^{offset}$, where $T_c^{onset}$ and $T_c^{offset}$ represent the onset and offset temperatures of the superconducting transition, respectively, as illustrated in Supplementary Figure Z7. These parameters provide complementary information on the evolution of the superconducting and normal-state transport properties with the applied pelletization pressure. As shown in Figure 5(c), the $T_c^{onset}$ increases from ~39 K for S1 sample to a maximum of ~44-45 K for S5, pelletized at 870 MPa. A further increase in the pelletization pressure results in a pronounced suppression of $T_c^{onset}$, demonstrating a clear non-monotonic pressure dependence with an optimum near 900 MPa. The increase in $T_c^{onset}$ up to the sample S5 coincides with an increase in the experimentally determined actual fluorine content $x_{act}$, a reduction of the lattice parameter $a$, and an increase in the relative sample density. These correlations suggest that changes in composition, phase assemblage, and densification may collectively contribute to the observed evolution of the superconducting transition. Furthermore, the enhancement of $T_c^{onset}$ across the pelletization-pressure series is also accompanied by a comparably small shift in the vibrational modes observed in Raman measurements. The subsequent decrease in $T_c^{onset}$ for S6 and S7 samples occurs together with an increase in secondary-phase fractions, and a reduction in the sample density (Supplementary Table ZT3). These correlations indicate that the superconducting response of these samples depends on the combined evolution of composition, phase assemblage, and bulk sample quality.

The transition width, $\Delta T$, exhibits a non-monotonic evolution across the pelletization-pressure series, as shown in Figure 5(d). For the sample S1, the superconducting transition is relatively broad, with $\Delta T$ of ~7 K. Increasing the pelletization pressure from S1 toward S5 is associated with a narrowing of the transition, reaching ~2.6 K for sample S5. The sharpening of the transition is consistent with a more uniform macroscopic superconducting response and may reflect reduced variations in local superconducting properties or improved effective intergrain coupling. This interpretation is further supported by the magnetic measurements, where the double-step feature observed in the ZFC magnetization curves at lower pelletization pressures becomes less pronounced toward the intermediate pelletization-pressure range. This change is consistent with an improved effective intergrain coupling. The simultaneous increase in the effective vortex-activation energy obtained from the thermally activated flux-flow analysis, as discussed below, further suggests that thermally activated vortex motion is more strongly suppressed in the favorable processing range. At higher pelletization pressures, the transition width no longer decreases and increases markedly for S7. The broadening of the transition coincides with the changes in the secondary-phase fractions identified by XRD and the reduction in the sample density. Although XRD and Raman spectroscopy indicate only modest changes in the average crystal structure and measured vibrational response, the changes in phase assemblage and porosity may contribute to a less uniform macroscopic superconducting transition. Thus, the evolution of $\Delta T$ provides additional evidence that S5, prepared at 870 MPa, exhibits the most favorable combination of transition temperature and transition sharpness among the investigated samples.

The residual resistivity ratio, *RRR*, provides complementary information on the normal-state transport properties, as shown in Figure 5(e). The *RRR* increases from S1 toward S5, indicating a change in the normal-state resistivity ratio over this part of the pelletization-pressure series. This evolution coincides with the increases in experimentally determined fluorine content $x_{act}$ and the sample density and may be associated with changes in the effective conducting pathways through the polycrystalline samples. At higher pelletization pressures, *RRR* decreases for S6 and S7, indicating a reversal in this trend. This decrease occurs alongside the increasing fraction of secondary phases and a reduction in the sample density, suggesting that increased phase inhomogeneity and changes in the effective conducting pathways may contribute to the observed normal-state transport behavior. Taken together, the evaluation of $T_c^{onset}$, $\Delta T$, and $RRR$ identifies S5, pelletized at 870 MPa, as the sample with the most favorable overall combination of superconducting transition temperature, transition width, and normal-state resistivity ratio. Up to this processing condition, increasing pelletization pressure is associated with a favorable combination of EDX-derived fluorine content $x_{act}$, final densification, $T_c$, superconducting transition width, and $RRR$. Further increasing the pelletization pressure (> 900 MPa) is associated with increased secondary-phase content and reduced the sample density, together with a lower transition temperature, a broader superconducting transition and degraded normal-state transport. These results demonstrate that the pelletization pressure can serve as an effective processing parameter for tailoring the superconducting and electrical transport properties of $PrFeAsO_{0.7}F_{0.3}$. Figure 5(c)-(e) is also shown in the supplementary Figure Z9 without guide-to-the-eye lines, and its dependence on the actual fluorine content $x_{act}$ derived from EDX is depicted in the supplementary Figure Z10. Furthermore, to assess sample reproducibility, the measured normalized resistance curves for two selected samples, S5 and S7, are presented in Supplementary Figure Z8.

Magnetotransport measurements were performed under the applied magnetic fields up to 9 T to investigate the field-dependent superconducting transitions and vortex-transport behaviour of $PrFeAsO_{0.7}F_{0.3}$ prepared under different pellettization pressure. Samples S5 and S6 were selected for detailed analysis because they exhibit relatively sharp superconducting transitions, with transition widths $\Delta T$ of ~2.6 K at zero magnetic field. The relatively narrow transitions allow the field-induced broadening of the superconducting transition to be evaluated with the improved reliability. The temperature dependence of the electrical resistivity was measured under applied magnetic fields up to 9 T, and the corresponding results for S5 and S6 are presented in Figures 6(a) and 6(b), respectively. For both samples, application of an external magnetic field progressively broadens the superconducting transition and shifts it toward lower temperatures. At 9 T, the superconducting transition widths increase to around 18.5 K and 17.3 K for the sample S5 and S6, respectively. This broadening behavior is characteristic of the resistive transition in type-II superconductors and is consistent with an increasing contribution from vortex motion to electrical dissipation in the mixed state. The upper critical field, $H_{c2}$, and irreversibility field, $H_{irr}$, were determined from resistive curves using criteria corresponding to 90% and 10%, respectively, of the normal-state resistivity, $\rho_{55K}$, measured above the superconducting transition, by following the Werthamer–Helfand–Hohenberg (WHH) model [36]: $H_{c_2}(0) = -0.693 \cdot \frac{dH_{c_2}}{dT}\Big|_{T=T_c} . T_c$. The resulting field-temperature dependences of $H_{c2}(T)$ and $H_{irr}(T)$ are shown in Figure 6(c). The initial slopes of the resistively determined upper-critical-field curves near the onset transition temperature, $T_c^{onset}$ , $\left.\frac{dHc_2}{dT}\right|_{T=T_c}$, are approximately -14.9 T K$^{-1}$ for S5 sample and -14.2 T K$^{-1}$ for S6 sample. These values are considerably larger than the previously reported slope magnitude of ~6.3 T K$^{-1}$ for $PrFeAsO_{0.7}F_{0.3}$ [6]. However, similar steep upper-critical-field slopes have also been reported for F doped Sm1111 single crystals [37]. Despite these steep initial slopes, the applied magnetic-field range of 9 T is insufficient to determine the zero-temperature upper critical field, $H_{c2}(0)$, reliably. Accordingly, the present analysis is restricted to the experimentally accessible field range, and no numerical WHH-derived value of $H_{c2}(0)$ is assigned for these samples. The difference between the measured initial slopes and the previously reported value may be associated with differences in composition, scattering, phase assemblage, or microstructural characteristics. However, the available measurements do not allow these possible contributions to be separated quantitatively. The measured slopes therefore describe the near- $T_c$ behaviour of the resistive transition and should not be taken as evidence for a particular microscopic mechanism or used alone to establish the magnitude of $H_{c2}(0)$. The initial slopes of the irreversibility-field $H_{irr}$ curves, expressed as the magnitudes $\left.\frac{dHirr}{dT}\right|_{T=T_c}$, are approximately -1.78 T K$^{-1}$ for S5 and -1.40 T K$^{-1}$ for S6. These values are comparable to or slightly larger in magnitude than the previously reported value of ~ -1.3 T K$^{-1}$ for $PrFeAsO_{0.7}F_{0.3}$ [6]. The separation between $H_{c2}$ and $H_{irr}$ curves identifies a field–temperature region in which superconducting fluctuations and vortex motion can contribute to finite electrical dissipation below the resistively defined upper-critical-field criterion. The lower $H_{irr}$ slope observed for S6 indicates a difference in the field dependence of its resistive irreversibility relative to S5. This difference is considered alongside the TAFF results below, although the two measurements do not independently identify its microscopic origin. The differences between S5 and S6, despite their relatively similar near $T_c$ upper-critical-field slopes, indicate that their resistive irreversibility and vortex-transport responses differ. Because the pelletization pressure was completely released before heat treatment, the observed differences in $H_{c2}$ and $H_{irr}$ between S5 and S6 reflect the properties of the final prepared samples using different pelletization pressures. They should not be interpreted as the direct effect of pressure acting during superconducting phase formation.

The thermally activated flux-flow (TAFF) analysis of samples S5 and S6 provides further insight into their field-dependent vortex-transport behaviour. The corresponding resistivity data and Arrhenius analysis are presented in Figures 7(a) and 7(b). In the mixed-state region, thermally activated vortex motion across effective energy barriers contributes to finite electrical resistivity. The corresponding activation energy, $U_0(H)$, therefore characterizes the effective energy barrier for vortex motion under the applied magnetic field and the present experimental conditions. The temperature dependence of the resistivity in the TAFF regime was analysed using the Arrhenius relation [28] [6, 17]: $\rho\ (T,H) = \rho_0(H)\ exp\left(-\frac{U_0(H)}{k_B T}\right)$, where $\rho_0(H)$ is a field-dependent prefactor, $k_B$ is the Boltzmann constant, and $U_0(H)$ is the effective vortex-activation energy. The activation energies were extracted from the slopes of the linear regions in plots of ln($\rho/\rho_{55K}$) versus $1/T$ as reported previously [6, 24, 32, 17]. The resulting $U_0$ ($H$) values for S5 and S6 samples are summarized in Figure 7(c). Sample S5 exhibits consistently higher $U_0(H)$ values than S6 over the investigated field range, suggesting a larger effective energy barrier for thermally activated vortex motion under the present experimental conditions. This difference occurs alongside the higher $T_c$, narrower zero-field transition, and more favorable effective magnetic critical-current performance observed for S5. However, the extracted activation energy may also be affected by transition broadening, $H_{irr}$, intergrain connectivity, phase assemblage, and microstructural characteristics. Therefore, the higher $U_0(H)$ of the sample S5 cannot be interpreted as direct evidence of stronger intrinsic vortex-pinning.

The field dependence of the effective activation energy was further analyzed using the power-law relationship, $U_0(H) \propto H^{-\eta}$, where $\eta$ characterizes the field dependence of the effective activation energy. For S5 sample, two field regimes are observed: at fields up to ~7 T, $\eta_1$~0.48, whereas at higher fields, $\eta_2$~0.79. For S6 sample, the corresponding exponents are $\eta_1$~0.42 and $\eta_2$~0.77. The change in slope of $U_0(H)$ indicates a crossover in its field dependence. At low fields, the effective activation energy may be particularly sensitive to sample-dependent factors, including current connectivity, microstructural heterogeneity, and weak-link behaviour. At higher fields, the increasing vortex density and vortex–vortex interactions may also contribute to the observed field dependence. Notably, the high-field exponents are very similar for S5 and S6 ($\eta_2$ ~0.79 and 0.77), indicating broadly comparable field dependence of the effective activation energy in this regime. The different low-field exponents, together with the consistently higher absolute $U_0(H)$ values for S5, indicate differences in effective vortex-transport behaviour but do not establish distinct intrinsic pinning mechanisms. The higher effective activation energy of S5 occurs alongside its higher relative density, higher EDX-derived actual fluorine content $x_{act}$, higher $T_c$, narrower transition, and more favorable magnetic critical-current performance, particularly at higher magnetic fields (as discussed below). Its comparatively compact microstructure and phase assemblage may favour macroscopic superconducting current transport and contribute to the observed activation-energy difference. In contrast, S6 exhibits lower relative density and a somewhat greater secondary-phase contribution, together with less favourable superconducting transport properties. Although secondary phases and structural defects can provide vortex-pinning sites under some conditions, microstructural heterogeneity may also disrupt current paths through a porous polycrystalline network [38]. Overall, S5 exhibits more favourable effective vortex-activation behaviour than S6 under the present measurement conditions. Nevertheless, the similar high-field exponents and the possible influence of transition broadening, irreversibility, and intergrain connectivity prevent the difference in $U_0(H)$ from being assigned uniquely to intrinsic pinning. Because the initial pelletization pressure was released before heat treatment, the observed differences reflect the properties of the final heat-treated samples prepared using different initial pelletization conditions.

## (e) Magnetic Properties

The magnetic response of samples S1–S7 was investigated using zero-field-cooled (ZFC) and field-cooled (FC) magnetization measurements under an applied magnetic field of 20 Oe, as shown in Figure 8(a). The magnetic superconducting transition temperature, $T_c^{mag}$, was determined from the onset of the diamagnetic response in the ZFC magnetization curves as similar to our previous reports [24, 6, 32, 19]. Samples S1 and S2 exhibit a pronounced two-step feature in their magnetic transitions, with $T_c^{mag}$ values of ~37.7 and 38.0 K, respectively. Such a two-step magnetic transition is commonly associated with distinct intragrain and intergrain superconducting responses and can be indicative of relatively weak intergranular coupling and/or superconducting inhomogeneity in polycrystalline superconductors [39, 40]. In the present samples, the observed feature may also reflect a distribution of local superconducting transition temperatures arising from compositional variations, secondary phases, or microstructural inhomogeneity [9]. The available ZFC/FC measurements do not allow these contributions to be separated unambiguously, particularly because no dedicated field-dependent analysis of the individual transition steps was performed [40]. With increasing pelletization pressure, the two-step feature becomes less pronounced and is absent in the predominantly single-step transitions observed for the samples S3–S7. This evolution indicates a change in the macroscopic magnetic response and may be consistent with changes in the effective intergrain connectivity and/or superconducting homogeneity [9, 39]. The relatively pronounced two-step transitions of S1 and S2 also coincide with their comparatively low measured relative sample densities. This correlation suggests that porosity and the associated limitations on macroscopic superconducting current paths may contribute to the observed behaviour, but it does not establish density as its sole cause.

The magnetic transition temperature increases with pelletization pressure up to S5 sample, reaching ~42.6 K for the sample compacted at 870 MPa, before decreasing at higher pressures. This trend is consistent with that observed in the electrical-resistivity measurements, indicating that the processing-dependent changes are reflected in both the magnetic and transport responses. The magnetic transition temperatures are ~1–2 K lower than the corresponding transport-derived $T_c$ values [6, 32]. Such a difference can arise because resistivity and magnetization probe different aspects of the superconducting response: electrical transport depends on the formation of a continuous superconducting current path, whereas magnetic measurements reflect the diamagnetic shielding response of the sample. As the ZFC/FC data were not used to determine a calibrated superconducting volume fraction of these polycrystalline samples, the magnitude of the diamagnetic response is discussed here comparatively rather than as a quantitative measure of the bulk superconducting fraction. The more clearly defined magnetic transition observed for S5 coincides with the maximum measured relative density in the sample series and with the predominance of the PrFeAs(O,F) phase identified by X-ray diffraction. The EDX-derived fluorine indicator also increases across much of the series, although it should be regarded as a comparative estimate rather than an absolute measure of bulk fluorine content. Thus, the favourable magnetic response of S5 correlates with changes in relative sample density, EDX-derived composition, and phase assemblage. In general, these magnetic measurements suggest that the improved response primarily reflects changes in superconducting composition i.e. potentially associated with increased fluorine incorporation into the Pr1111 lattice.

At higher pelletization pressures, the magnetic response deteriorates. For S6, the superconducting transition remains relatively well defined but occurs at a lower temperature than for S5. The suppression and broadening of the transition are more pronounced for S7, prepared at 1131 MPa. These changes coincide with a reduction in measured relative sample density compared with S5 and an increased contribution from secondary phases, particularly for S7. The concurrent evolution may be associated with increased superconducting inhomogeneity and/or less effective macroscopic current connectivity, however, the present measurements do not establish a unique mechanism for the deterioration at the highest pelletization pressure. Therefore, increasing the initial pelletization pressure beyond the intermediate-pressure condition does not produce a further improvement in the magnetic response under CSP-AP used here. The magnetic results are also consistent with the TAFF analysis, in which S5 exhibits a higher effective vortex-activation energy than S6. This finding indicates a higher effective energy barrier for thermally activated vortex motion under the measurement conditions. However, the extracted activation energy may be influenced not only by vortex pinning but potentially also by transition broadening, phase assemblage, microstructural heterogeneity, and intergrain connectivity. The magnetic and TAFF results provide complementary evidence of the more favourable overall superconducting response of S5. Overall, the magnetic measurements show that increasing the pelletization pressure from 174 MPa to 870 MPa is associated with a higher $T_c^{mag}$, a less pronounced two-step transition, and a more clearly defined macroscopic magnetic response. At higher pelletization pressures, the magnetic response deteriorates alongside changes in relative sample density and phase assemblage. Since the applied pressure was released before heat treatment, these observations describe differences between the final samples prepared using different initial pelletization pressures; they do not demonstrate a direct pressure effect during phase formation. Together with the transport, structural, compositional, density, and microstructural results, the magnetic measurements identify 870 MPa as the most favourable pelletization pressure among the conditions investigated for various $PrFeAsO_{0.7}F_{0.3}$ samples under the fixed CSP-AP schedule.

Magnetic hysteresis (*M*–*H*) loops were measured for samples S1–S7 to evaluate their effective magnetic current-carrying capability and field-dependent superconducting response. The critical current density, $J_c$, was estimated from the width of the magnetization hysteresis loops using the Bean critical-state model [27]: $J_c = 20\Delta M / V[a(1 - a/3b)]$, where $\Delta M$ is the difference between the magnetization values of the increasing- and decreasing-field branches, $V$ is the sample volume and, $a$ and $b$ ($a < b$) are the sample dimensions perpendicular to the applied magnetic field. The *M*–*H* loops measured at 5 K were corrected by subtracting the corresponding 60 K loops to reduce the normal-state magnetic background, as described in the Experimental section and illustrated in Supplementary Figure Z12. A residual field-dependent contribution remains after subtraction and may reflect temperature-dependent magnetic contributions associated with $Pr^{3+}$. The resulting hysteresis behaviour is consistent with previous reports on F-doped PrFeAs(O,F) superconductors [41, 29]. The resulting field dependence of $J_c$ for samples S1–S7 is presented in Figure 8(b). Because the samples are porous polycrystalline materials, the Bean-model values are treated as effective magnetic critical current densities, reflecting the macroscopic current-carrying response of the samples rather than intrinsic transport $J_c$ values for fully dense material as previously reported for IBS [6, 24, 32, 21, 41]. Samples S1 and S2 exhibit relatively low $J_c$ values, of the order of $10^2$ A cm$^{-2}$ across the investigated magnetic-field range. The sample S3 exhibits a moderate enhancement of $J_c$, particularly in the low-field region below ~3 T, but its $J_c$ gradually approaches values comparable to those of S1 and S2 at higher fields. Sample S4 shows a $J_c(H)$ dependence broadly similar to that of the lower-pressure samples. In contrast, S5, pelletized at 870 MPa, exhibits the most favourable $J_c(H)$ performance over most of the investigated field range, particularly at higher fields, with the maximum values of the order of ~$10^3$ A-cm$^{-2}$. This is approximately an order of magnitude higher than the values observed for S1 and S2, based on the reported orders of magnitude, and is comparable to previously reported values for $PrFeAsO_{0.7}F_{0.3}$ [6, 41]. This comparison concerns the effective magnetic $J_c$ and should not be interpreted as evidence of a corresponding increase in intrinsic vortex-pinning strength. The enhanced effective $J_c$ of S5 coincides with its highest measured relative sample density, higher superconducting transition temperature, and narrower transition. S5 also exhibits a higher effective vortex-activation energy than S6 in the TAFF analysis. Together, these observations are consistent with a more favourable macroscopic superconducting response in S5. However, the effective magnetic $J_c$ of a polycrystalline material depends on the current-carrying network and may be influenced by porosity, intergrain connectivity, phase assemblage, and microstructural inhomogeneity. The enhanced effective magnetic $J_c$ observed for S5 may be associated with its improved macroscopic current-carrying network, which may reflect its higher relative density, better intergrain connectivity, and more favourable microstructural characteristics, rather than stronger intrinsic vortex pinning. The field dependence of $J_c$ further illustrates the differences among the samples. S6 exhibits values comparable to those of S5 in the low-field region, but its $J_c$ decreases more rapidly as the magnetic field increases. Consequently, the effective current-carrying performance of S6 is less favourable than that of S5 at higher fields. This behaviour is consistent with the lower effective vortex-activation energy observed for S6, although the TAFF and magnetic hysteresis results do not establish a unique microscopic explanation for the different field dependences.

At the highest pelletization pressure, S7 sample, prepared at ∼1131 MPa, exhibits strongly suppressed $J_c$ values across the investigated field range, approaching those of S1 and S2 samples. This deterioration coincides with a lower superconducting transition temperature, a broader magnetic transition, reduced relative density compared with S5, and an increased contribution from secondary phases, particularly PrOF. Non-superconducting secondary phases and pores may interrupt effective superconducting current paths, while compositional and microstructural inhomogeneity may further limit the macroscopic current-carrying response. Although defects and secondary phases can also act as vortex-pinning sites, the strongly reduced $J_c$ of S7 suggests that the deterioration of the effective superconducting current-carrying network dominates over any possible beneficial contribution from additional defect- or secondary-phase-related pinning. Hence, the $J_c(H)$ results identify S5, prepared using a pelletization pressure of 870 MPa, as the most favourable sample among those investigated, particularly in the higher-field region. Increasing the pelletization pressure from 174 MPa to 870 MPa is associated with increased relative density, a less pronounced two-step magnetic transition, and enhanced effective magnetic $J_c$. At higher pelletization pressures, reduced relative density, changes in phase assemblage, and deterioration of the superconducting transition coincide with less favourable current-carrying performance. These correlations indicate that the effective magnetic $J_c$ reflects the combined influence of the final sample structure, phase composition, porosity, and macroscopic current connectivity. Taken together, the $J_c(H)$ results demonstrate that the initial pelletization condition is a relevant processing parameter for tailoring the effective macroscopic superconducting current-carrying capability of $PrFeAsO_{0.7}F_{0.3}$. The sample pelletized at 870 MPa exhibits the most favorable combination of high $J_c$, high $T_c$, the highest relative sample density, and relatively large effective vortex-activation energy, particularly in the higher-field region.

## (f) Discussion

Figures 9(a)-(c) summarize the evolution of the lattice parameter $a$, superconducting transition temperatures obtained from electrical resistivity ($T_c^{onset}$) and magnetization ($T_c^{mag}$), and effective magnetic $J_c$ at 0.5 T and 6 T at 5 K. The lattice parameter $a$ decreases with increasing pelletization pressure up to the intermediate-pressure range, reaching its minimum around samples S5–S6, while the EDX-derived actual fluorine content $x_{act}$ increases (Figure 2(f)). The corresponding variations in $a$, $c$, and $V$ as functions of $x_{act}$, together with comparisons with previously reported data [6], are presented in Supplementary Figures Z3(a)-(b), and Z4. Because fluorine substitution for oxygen is associated with lattice contraction in PrFeAsO-based compounds [6, 7, 8], the observed trends are qualitatively consistent with changes in fluorine incorporation across the sample series. However, the structural changes are relatively small, and XRD does not independently establish the fluorine occupancy of the superconducting phase. The comparatively modest variations in the Raman-active phonon frequencies are consistent with these limited structural changes but do not independently establish the extent of fluorine incorporation within the Pr1111 lattice.

The superconducting transition temperatures increase with pelletization pressure up to sample S5 and subsequently decrease at higher pressures (Figure 9(b)). Sample S5, compacted at 870 MPa, exhibits the highest $T_c^{onset}$ and $T_c^{mag}$ within the investigated series. At lower pelletization pressures, broader resistive transitions and more pronounced two-step magnetic transitions are observed. These features may reflect superconducting inhomogeneity, less effective intergranular coupling or a combination of these effects [40, 39, 4]. The increase in final sample density and the weakening of the two-step magnetic response toward S5 are consistent with improved macroscopic superconducting connectivity and/or a more uniform superconducting response. Nevertheless, because sample density, microstructure, phase composition, and the EDX-derived actual fluorine content vary together, their individual contributions to the evolution of $T_c$ cannot be separated in the present study. The behavior changes at higher pelletization pressures. For S6 and, more prominently, S7, the lattice parameter $a$ and unit-cell volume $V$ increase despite the continued increase or near-saturation of the overall EDX-derived fluorine content $x_{act}$. This behavior coincides with a decrease in $T_c$ and an increased contribution from the PrOF secondary phase, particularly in the sample S7 (Supplementary Figure Z1). The combined observations suggest that the overall fluorine content measured by EDX does not necessarily reflect the amount of fluorine incorporated into the superconducting PrFeAs(O,F) phase. A redistribution of fluorine between the superconducting Pr1111 phase and PrOF may contribute to the observed trends, especially for S7; however, the present measurements do not directly determine fluorine partitioning between these phases. The increase in the FeAs secondary phase and the changes in the final microstructure may also contribute to the deterioration of the superconducting response at the highest pelletization pressure.

To examine these trends further, the EDX-derived fluorine content $x_{act}$ was compared with the structural and superconducting properties of the individual samples. As shown in Supplementary Figure Z1, the estimated PrOF phase contribution remains approximately constant for S1–S6 and increases noticeably for S7. However, EDX measurements of fluorine $x_{act}$ are subject to limitations associated with the low-energy F Kα signal, spectral overlap, and sample-dependent effects such as porosity and surface or matrix conditions. Consequently, $x_{act}$ should be interpreted as an average over the analyzed sample and may include fluorine associated with both the PrFeAs(O,F) phase and the PrOF secondary phase. It therefore cannot be interpreted directly as the fluorine concentration within the superconducting Pr1111 phase alone. For S1 and S2, $x_{act}$ changes only slightly, while small changes in $T_c$ and the lattice parameter $a$ are also observed (Supplementary Figures Z3 and Z14(c)). From S3 to S5, an increase in $x_{act}$ accompanies a decrease in $a$ and an increase in $T_c$. The approximately constant PrOF contribution over this range is consistent with, but does not demonstrate, increased fluorine incorporation into the superconducting PrFeAs(O,F) phase. For S6 and S7, the increase or near-saturation of $x_{act}$ (Figure 2(f)), together with the reversal in the lattice-parameter trend (Supplementary Figures Z4 and Z3), indicates that the relationship between the overall EDX-derived fluorine content $x_{act}$ and the superconducting properties is related to each other. This behaviour is consistent with the complex composition–structure–property relationships reported for F-doped *RE*1111 compounds [25, 42]. These combined observations may reflect changes in fluorine distribution between the superconducting and secondary phases, as well as differences in phase assemblage and microstructure.

Comparisons with previously reported results [6, 7] provide additional context. Supplementary Figure Z14(b) presents the EDX-derived fluorine content $x_{act}$ as a function of nominal fluorine content ($x$), alongside data reported by Priya *et al.* [6] and Rotundu *et al.* [7]. Because the nominal composition is identical for S1–S7, this comparison provides a reference for assessing the range of the actual fluorine contents rather than an independent measure of pressure-dependent incorporation. Samples S5 and S6 approach the trends established by the previously reported these data. Supplementary Figures Z14(c) and Z13(b) compare $T_c$ with $x_{act}$, while Supplementary Figures Z4, Z3 and Z13(a) compare the lattice parameter and unit-cell volume with $x_{act}$. From S1 to S5, the increase in $x_{act}$ accompanies a decrease in $a$ and $V$, and an increase in $T_c$, broadly consistent with the trends reported in Ref. [7, 6]. Beyond S5, the increase in $a$ and $V$, and the decrease in $T_c$, despite the comparatively high EDX-derived fluorine content, indicate a departure from this relationship. These comparisons establish correlations among the actual fluorine content $x_{act}$, structural parameters, and superconducting properties, but they do not independently establish the absolute fluorine content, its occupancy of the superconducting phase, or its distribution between phases. Nevertheless, the observed trends suggest that the enhancement of the superconducting transition temperature may be associated with increased fluorine incorporation into the superconducting Pr1111 lattice, consistent with previous reports [7, 6].

Figure 9(c) shows the field-dependent effective magnetic critical current density $J_c$ at 5 K. At 0.5 T, the highest $J_c$ is obtained for S6, compacted at ~1044 MPa, whereas at 6 T, S5 exhibits the highest $J_c$. Thus, the pelletization condition associated with the highest low-field current density is not necessarily the one providing the best performance at higher magnetic fields. The relatively high low-field $J_c$ of S6 may reflect its effective superconducting connectivity, microstructure, phase assemblage, and vortex dynamics. However, its more rapid decrease in $J_c$ with increasing field indicates that this advantage is not maintained at higher fields. In contrast, S5 exhibits a more favorable field dependence and retains a comparatively high $J_c$ at 6 T. This behavior is consistent with the larger thermally activated flux-flow (TAFF) activation energy observed for S5, indicating a higher effective energy barrier against thermally activated vortex motion under the investigated conditions [28]. The TAFF activation energy is an effective quantity that can also depend on transition broadening, irreversibility, microstructure, and intergranular connectivity; it should therefore not be interpreted as a direct measure of intrinsic vortex pinning alone. Similarly, the Bean-model values reported here represent effective magnetic critical current densities for porous polycrystalline samples rather than intrinsic or transport critical current densities. Hence, these results identify S5 as the most favourable sample overall among the investigated conditions, although S6 exhibits the highest low-field $J_c$. S5 combines the highest superconducting transition temperatures with the high-field $J_c$, a relatively narrow transition, the maximum relative sample density, and a larger effective TAFF activation energy. The structural, compositional, and superconducting trends suggest that the intermediate pelletization-pressure range is associated with a favorable combination of the final sample density, phase assemblage, the actual fluorine content, and effective superconducting connectivity. At lower pelletization pressures, lower EDX-derived actual fluorine contents, lower final densities, and greater porosity coincide with less favorable superconducting properties. At higher pressures (> 1000 MPa), the increase in secondary-phase contributions and reversal of the lattice-parameter trend coincide with deterioration of the superconducting response.

Importantly, the applied pressure in this study was used only for cold pelletization and was completely released before the common CSP-AP schedule. It therefore did not act as an externally applied pressure during phase formation. Rather, the initial pelletization condition was associated with subsequent differences in the final composition indicators, phase assemblage, density, porosity, microstructure, and superconducting properties after CSP-AP under otherwise identical conditions. The present results consequently establish a processing–structure–composition–property correlation, rather than an intrinsic pressure effect during phase formation. Although the combined evidence suggests that changes in fluorine incorporation may contribute to the observed evolution of the superconducting properties, the present EDX measurements cannot independently establish this contribution. Further work using a validated bulk fluorine-quantification method, phase-specific compositional analysis, and replicated synthesis across the pelletization-pressure range with different fluorine-doped Pr1111 compositions would be required to distinguish compositional effects from those associated with density, phase assemblage, and microstructure.

# Conclusion

This study investigated the influence of pelletization pressure ranging from 174 to 1131 MPa on the structural, microstructural, and superconducting properties of $PrFeAsO_{0.7}F_{0.3}$ prepared using CSP-AP. The applied pressure was released before the common synthesis procedure. The lattice parameter $a$ and unit-cell volume $V$ decreased with increasing pelletization pressure up to ~900 MPa and increased at higher pressures. The EDX-derived actual fluorine-content also increased across the sample series; however, Raman spectroscopy revealed only modest variations in the observed vibrational modes, without a pronounced systematic trend. Among the investigated samples, S5, pelletized at 870 MPa, exhibited the most favorable overall superconducting properties, including a superconducting transition temperature of 44.2 K, a narrower transition, the highest relative sample density, and a favorable effective magnetic critical-current density at high fields. At higher pelletization pressures, the increased secondary-phase contributions, the reversal in the lattice-parameter and unit-cell-volume trends, and the reduced sample density coincided with deterioration of the superconducting response. The observed behavior reflects the potentially coupled influences of composition, phase assemblage, porosity, and grain connectivity. Nevertheless, the observed correlations suggest that increased fluorine incorporation into the superconducting Pr1111 lattice may have contributed to the enhancement of the superconducting transition temperature, alongside changes in relative density, phase assemblage, and microstructure. Since the applied pressure was released before heat treatment, the results demonstrated processing–structure–composition–property correlations rather than an intrinsic pressure effect during phase formation. Thus, a pelletization pressure of 870 MPa represents the most favorable pelletization pressure within the investigated pressure range and under the employed CSP-AP schedule. Overall, this study highlights pelletization pressure as an important processing parameter for tailoring the superconducting properties of polycrystalline PrFeAs(O,F), with potential relevance to the processing of other iron-based superconductors using CSP-AP. Further studies are needed in this direction with different fluorine contents to clarify the mechanisms underlying these trends.

## CRediT authorship contribution statement

**Jayant Kumar:** Writing – review & editing, Writing – original draft, Investigation, Formal analysis, Validation, Data curation. **Konrad Kwatek:** Formal analysis, Investigation, Resources, Data curation, Writing – review & editing. **Tatiana Zajarniuk:** Data curation, Investigation, Resources, Writing – review & editing. **Taras Palasyuk:** Writing – review & editing, Formal analysis, Investigation, Data curation. **Cezariusz Jastrzębski:** Writing – review & editing, Resources, Data curation. **Tomasz Cetner:** Data curation, Resources. **Yaroslaw Konopelnyk:** Resources, Data curation. **Shiv J. Singh:** Writing–review & editing, Writing – original draft, Visualization, Validation, Supervision, Software, Resources, Methodology, Investigation, Funding acquisition, Formal analysis, Conceptualization.

## Declaration of competing interest

The authors declare that they have no known competing financial interests or personal relationships that could have appeared to influence the work reported in this paper.

## Data availability

The raw/processed data required to reproduce these findings cannot be shared at this time due to technical or time limitations. Data are available upon request to the corresponding author.

**Acknowledgments:**

This work was funded by Weave-UNISONO project (2025/07/Y/ST5/00116) sponsored by National Science Centre (NCN), Poland. SJS acknowledges financial support from National Science Centre (NCN), Poland through research Project number: 2025/07/Y/ST5/00116. We would also like to thank Priya Singh and M. Azam from the Institute of High Pressure Physics, Warsaw, Poland, for their help with some of the measurements and data analysis.

**Figure 1:** Pelletization pressure dependence of the relative sample density of $PrFeAsO_{0.7}F_{0.3}$ samples, as marked as S1 to S7, prepared using CSP-AP. The dashed curve is a guide to the observed pressure-dependent trend.

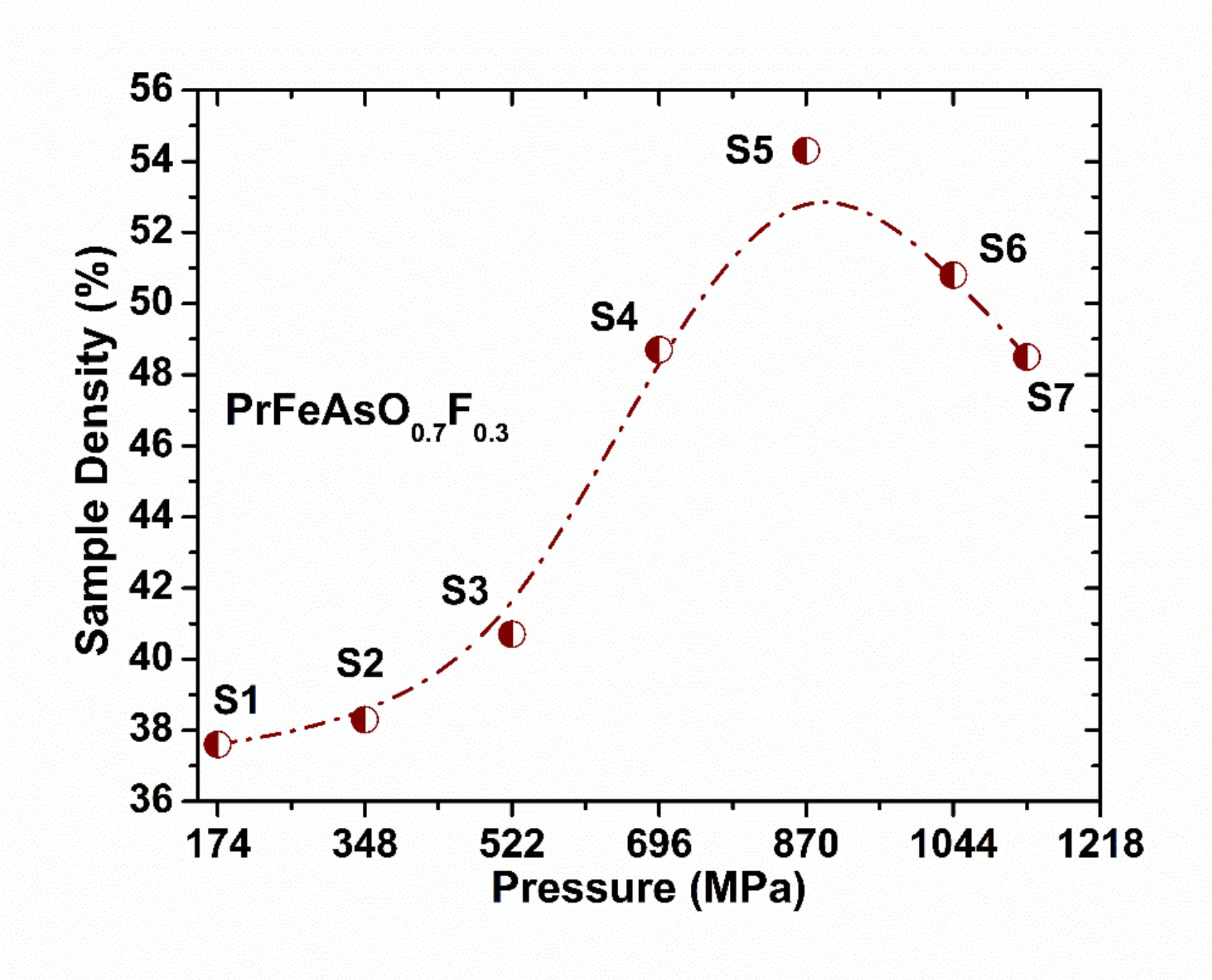

**Figure 2:** **(a)** Powder X-ray diffraction (XRD) patterns of $PrFeAsO_{0.7}F_{0.3}$ samples prepared under different pelletization pressures. **(b)** Rietveld refinement of the powder X-ray diffraction pattern for sample S5 is shown. The observed, calculated, and background profiles are shown by the corresponding symbols and lines, while the difference curve is displayed at the bottom. The vertical tick marks indicate the Bragg reflection positions of the PrFeAs(O,F) main phase, and the PrOF and FeAs secondary phases. The inset shows the crystal structure of the PrFeAs(O,F) (Pr1111) phase. Pelletization pressure dependence of the refined lattice parameters of $PrFeAsO_{0.7}F_{0.3}$: **(c)** *in-plane* lattice parameter $a$, **(d)** lattice parameter $c$, and **(e)** unit-cell volume $V$. The symbols represent the refined values with error bars, while the dashed curves are guides to the observed pressure-dependent trends. **(f)** Pelletization pressure dependence of the actual fluorine content, $x_{act}$, for the samples S1 to S7, obtained from EDX analysis. The dashed line is a guide to the observed trend, while the error bars represent the uncertainty in the EDX-derived composition. The corresponding plots (c)-(e) are provided in the Supplementary Information as Figure Z5 without guide-to-the-eye lines and Figure Z4 with the EDX-derived fluorine content estimate, $x_{act}$.

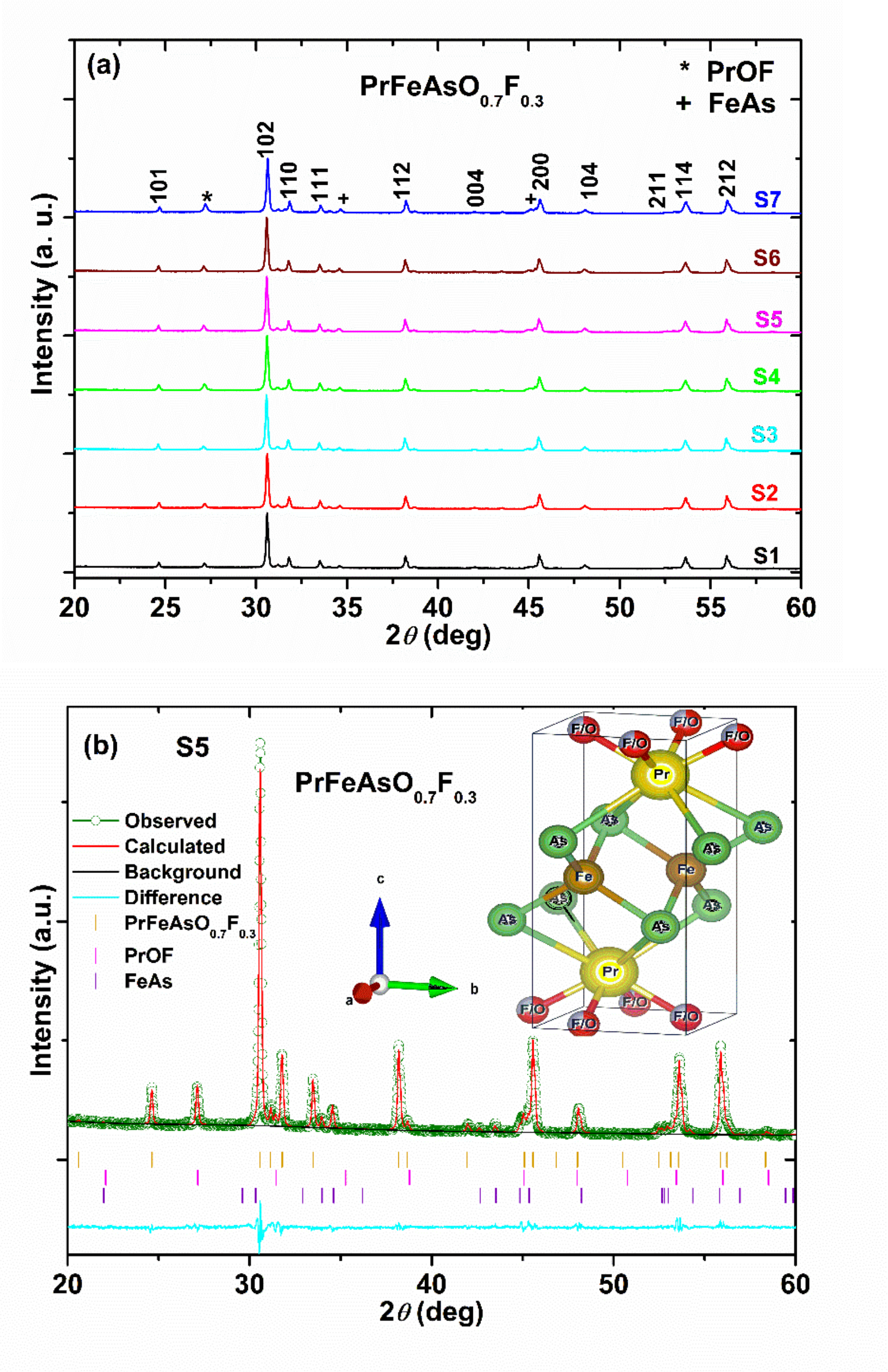

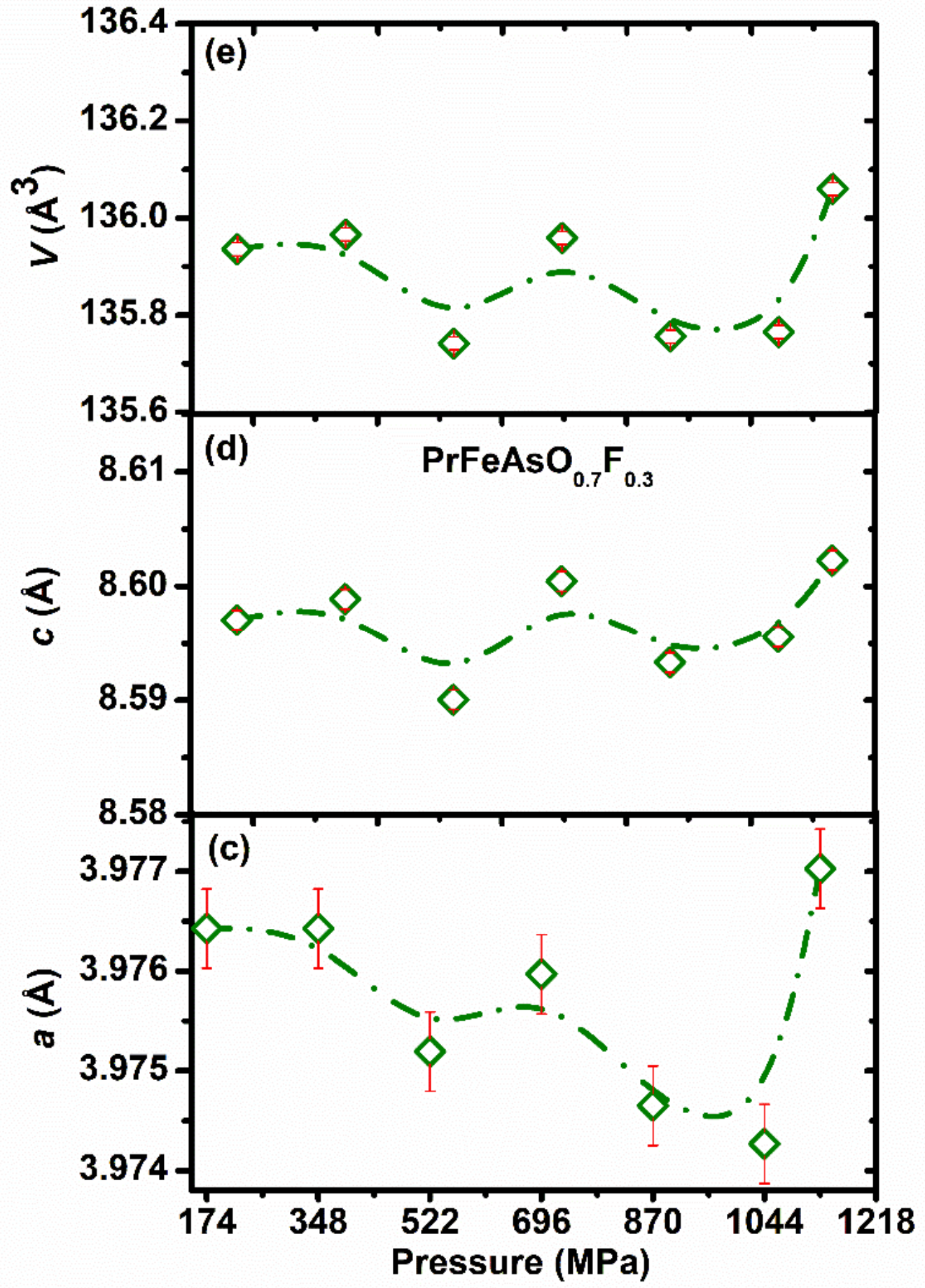

136.4
136.2
136.0
135.8
135.6
(e)
V (Å3)
(d)
PrFeAsO0.7F0.3
8.61
8.60
8.59
8.58
c (Å)
(c)
3.977
3.976
3.975
3.974
a (Å)
174
348
522
696
870
1044
1218
Pressure (MPa)

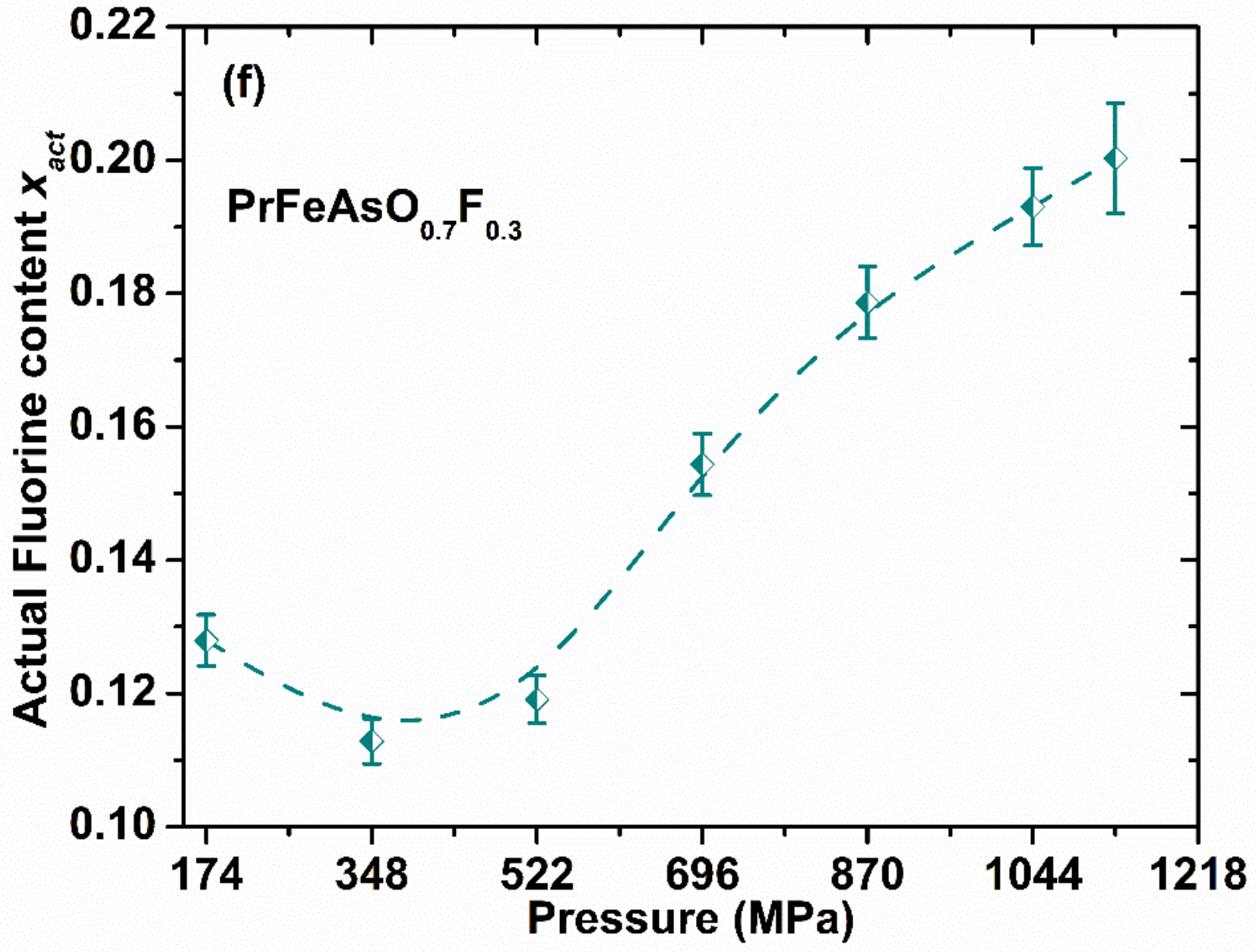

0.22
0.20
0.18
0.16
0.14
0.12
0.10
(f)
PrFeAsO0.7F0.3
Actual Fluorine content x act
174
348
522
696
870
1044
1218
Pressure (MPa)

**Figure 3:** Backscattered-electron (BSE) SEM micrographs of $PrFeAsO_{0.7}F_{0.3}$ samples prepared at different pelletization pressures: **(a–c)** S1, **(d–f)** S5, and **(g–i)** S7. Images are shown at progressively higher magnifications to illustrate the microstructural features. Bright, light-grey, and black contrasts correspond to $Pr_2O_3$ (or PrOF), the $PrFeAsO_{0.7}F_{0.3}$ main phase, and pores, respectively. Black-contrast regions may occasionally correspond to FeAs phase.

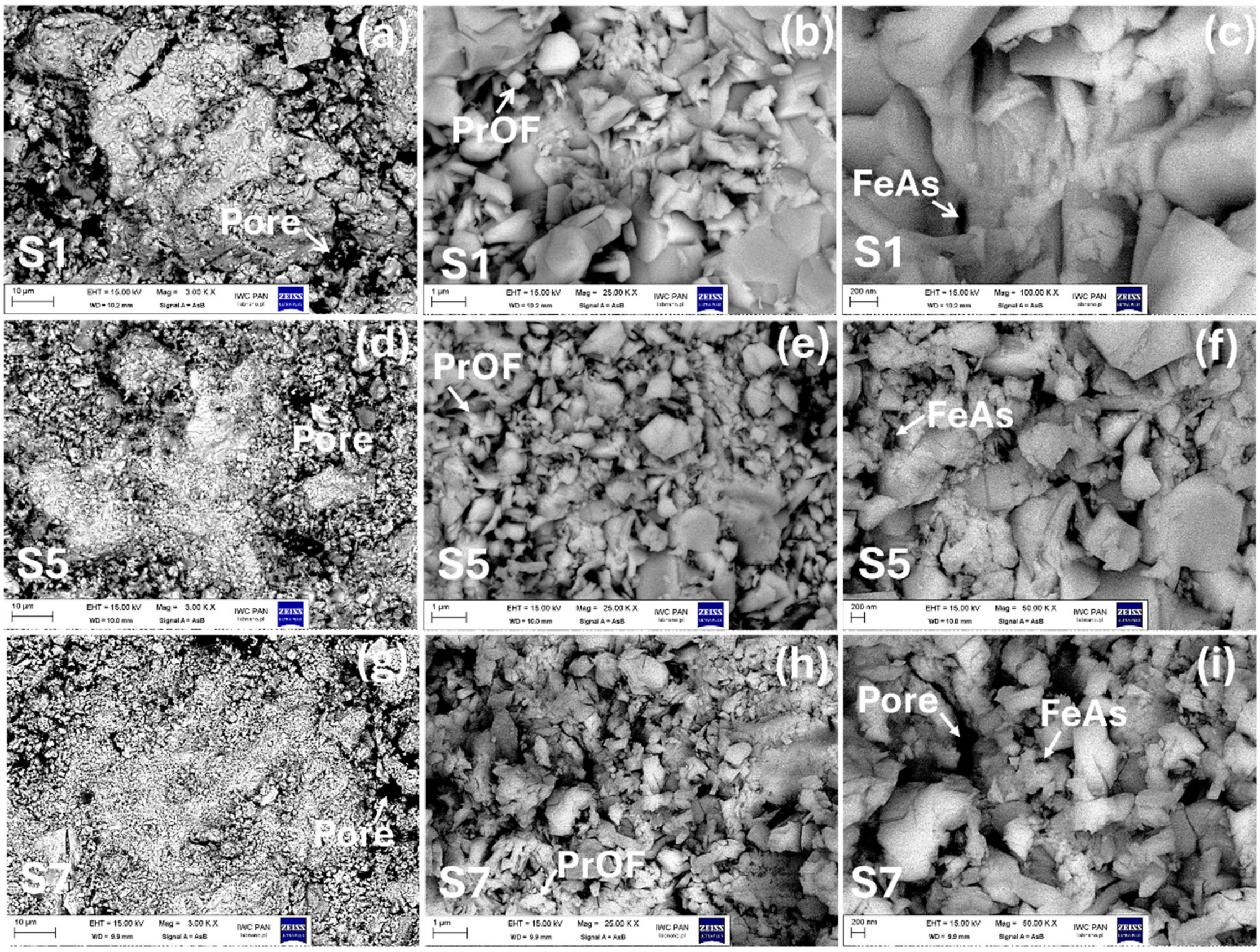

**Figure 4:** **(a)** Raman spectra of $PrFeAsO_{0.7}F_{0.3}$ samples S1–S7 prepared using different pelletization pressures (left panels). The experimental spectra (symbols) are fitted using Lorentzian components (green curves), with the resulting total fits shown by red curves. **(b)** The corresponding pelletization-pressure dependence of the Raman peak positions of the characteristic Raman-active phonon modes is shown in the right panels. The Fe $B_{1g}$ and As $A_{1g}$ modes exhibit a slight overall increase in frequency with increasing pelletization pressure, whereas the Pr $A_{1g}$ mode shows a more non-monotonic variation.

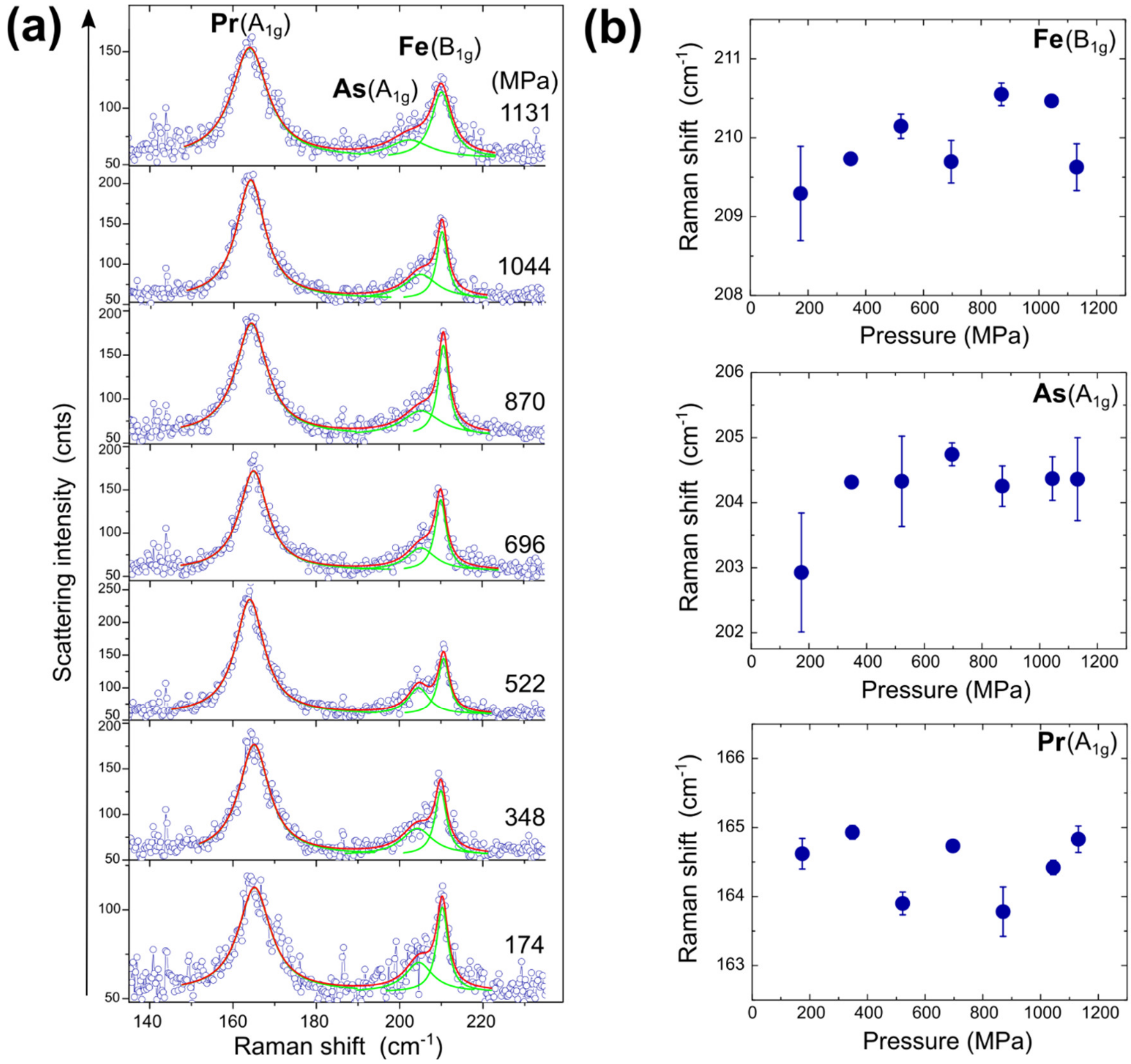

**Figure 5:** Temperature dependence of the electrical resistivity of $PrFeAsO_{0.7}F_{0.3}$ samples S1–S7 prepared using different pelletization pressures. **(a)** Resistivity over the full temperature range (up to ~280-290 K), showing the metallic normal-state behavior and superconducting transitions. **(b)** Expanded view of the superconducting transition region, highlighting the evolution of $T_c^{onset}$ and transition width with the pelletization pressure. Pelletization-pressure dependence of the superconducting and normal-state transport parameters of $PrFeAsO_{0.7}F_{0.3}$: **(c)** the onset superconducting transition temperature, $T_c^{onset}$, **(d)** the superconducting transition width, $\Delta T$, and **(e)** the residual resistivity ratio, $RRR = \rho_{280K}/\rho_{50K}$. The dashed curves are guides to the observed pelletization-pressure-dependent trends. The corresponding plots (c)-(e) are provided in the Supplementary Information as Figure Z9 without guide-to-the-eye lines and Figure Z10 with the EDX-derived actual fluorine content, $x_{act}$.

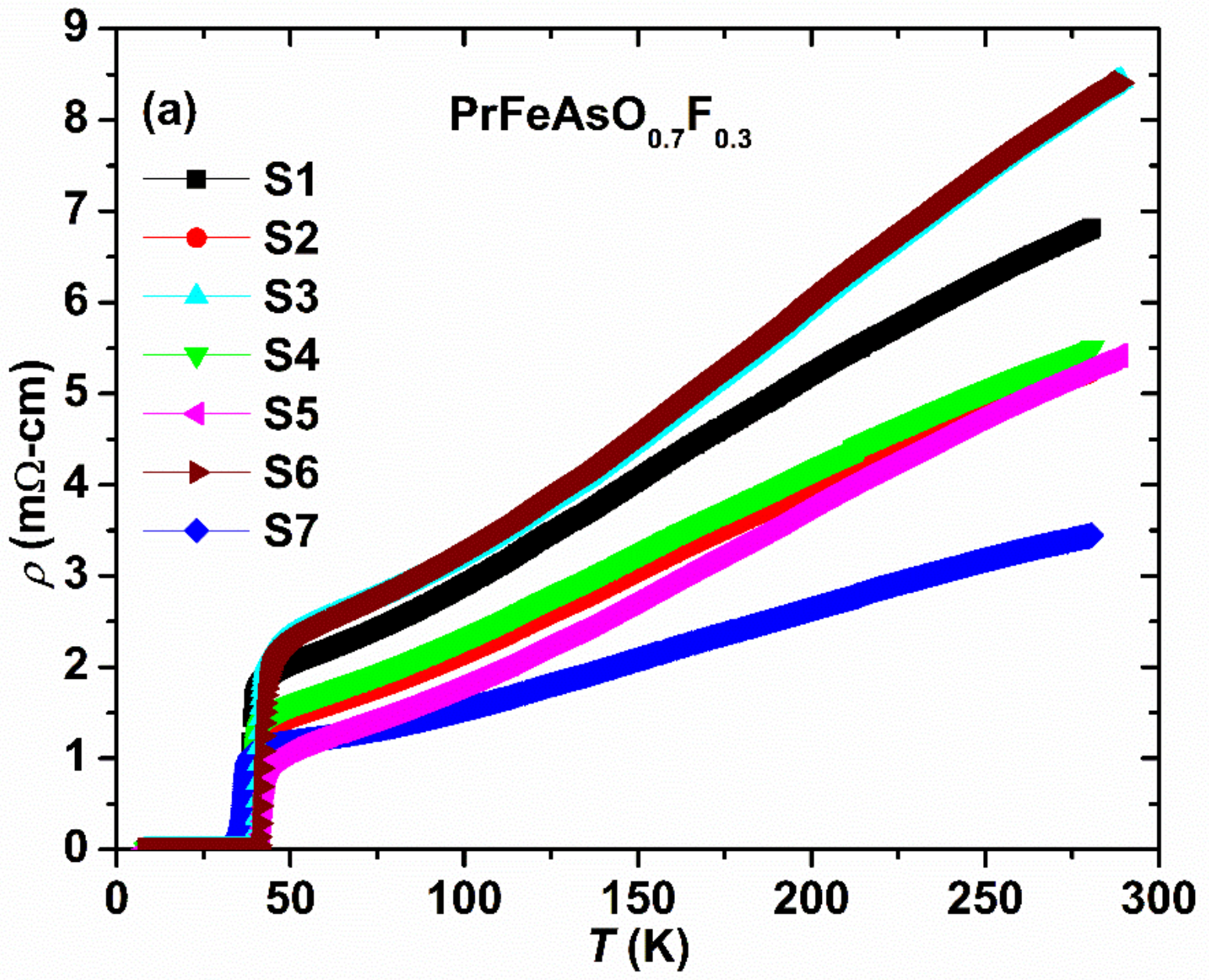

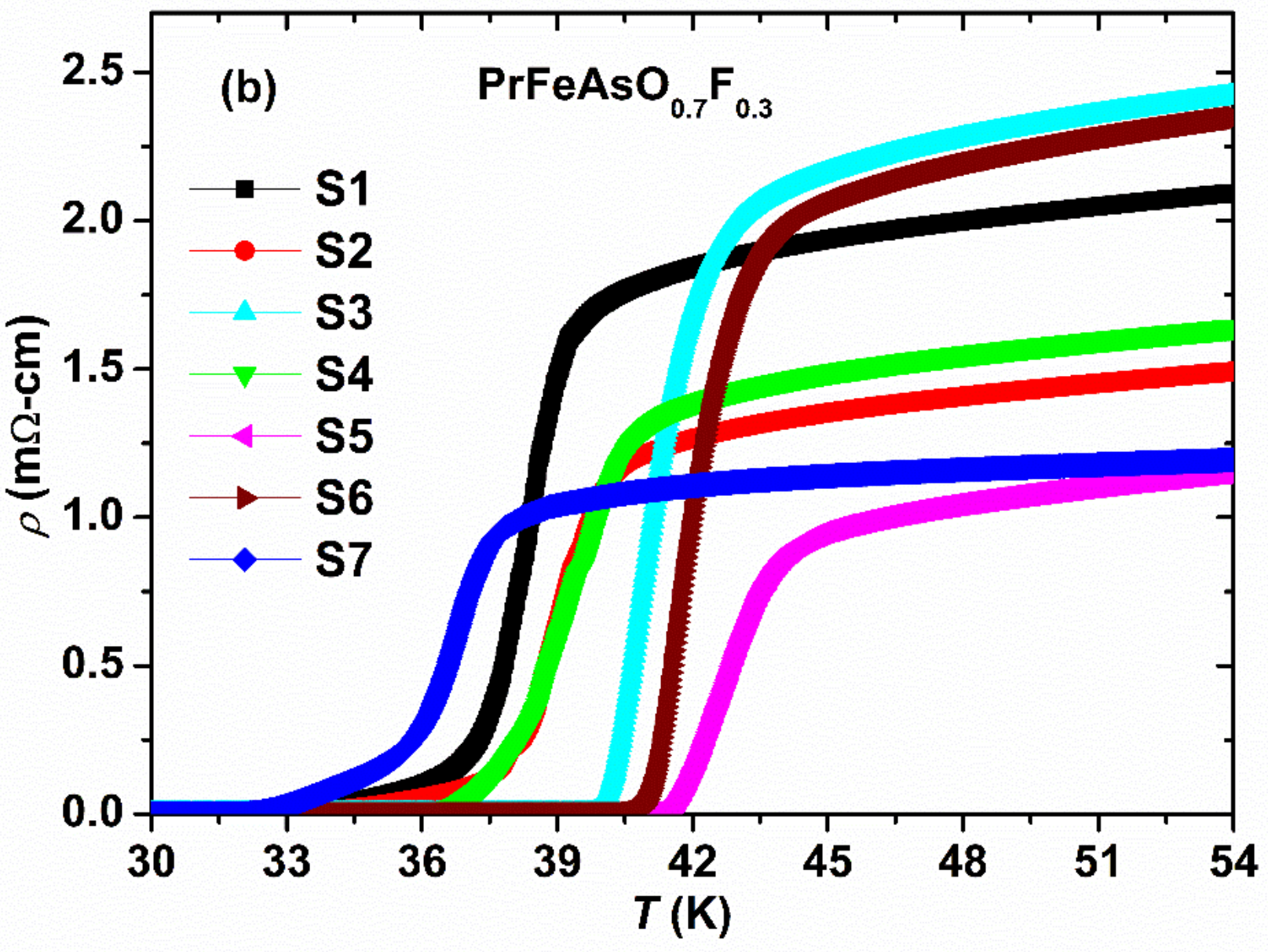

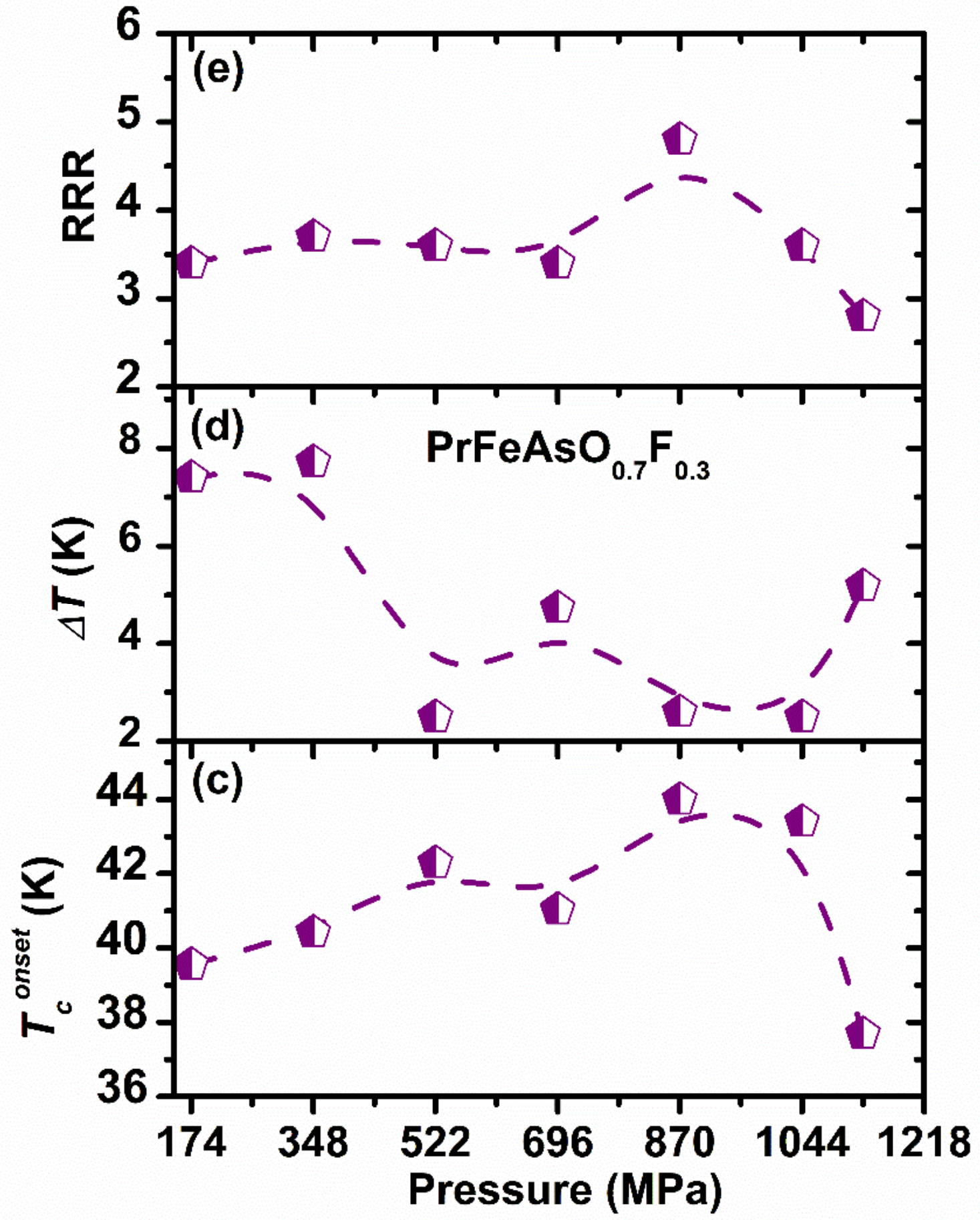

(e)
RRR
(d)
PrFeAsO0.7F0.3
ΔT (K)
(c)
Tc onset (K)
174
348
522
696
870
1044
1218
Pressure (MPa)

**Figure 6:** Temperature dependence of the resistivity of $PrFeAsO_{0.7}F_{0.3}$ under applied magnetic fields for samples S5 and S6. **(a)** and **(b)** Normalized resistivity, $\rho/\rho_{55K}$, measured at magnetic fields up to 9 T for S5 and S6 samples, respectively, showing the progressive broadening and downward shift of the superconducting transition with increasing magnetic field. **(c)** Temperature dependence of the upper critical field, $H_{c2}$, and irreversibility field, $H_{irr}$, determined using the 90% and 10% resistivity criteria, respectively. The dashed lines represent linear fits to the near $T_c$ regions, yielding the slope $dH_{c2}/dT|_{T=Tc}$ , and $dH_{irr}/dT|_{T=Tc}$ for S5 and S6 samples, respectively.

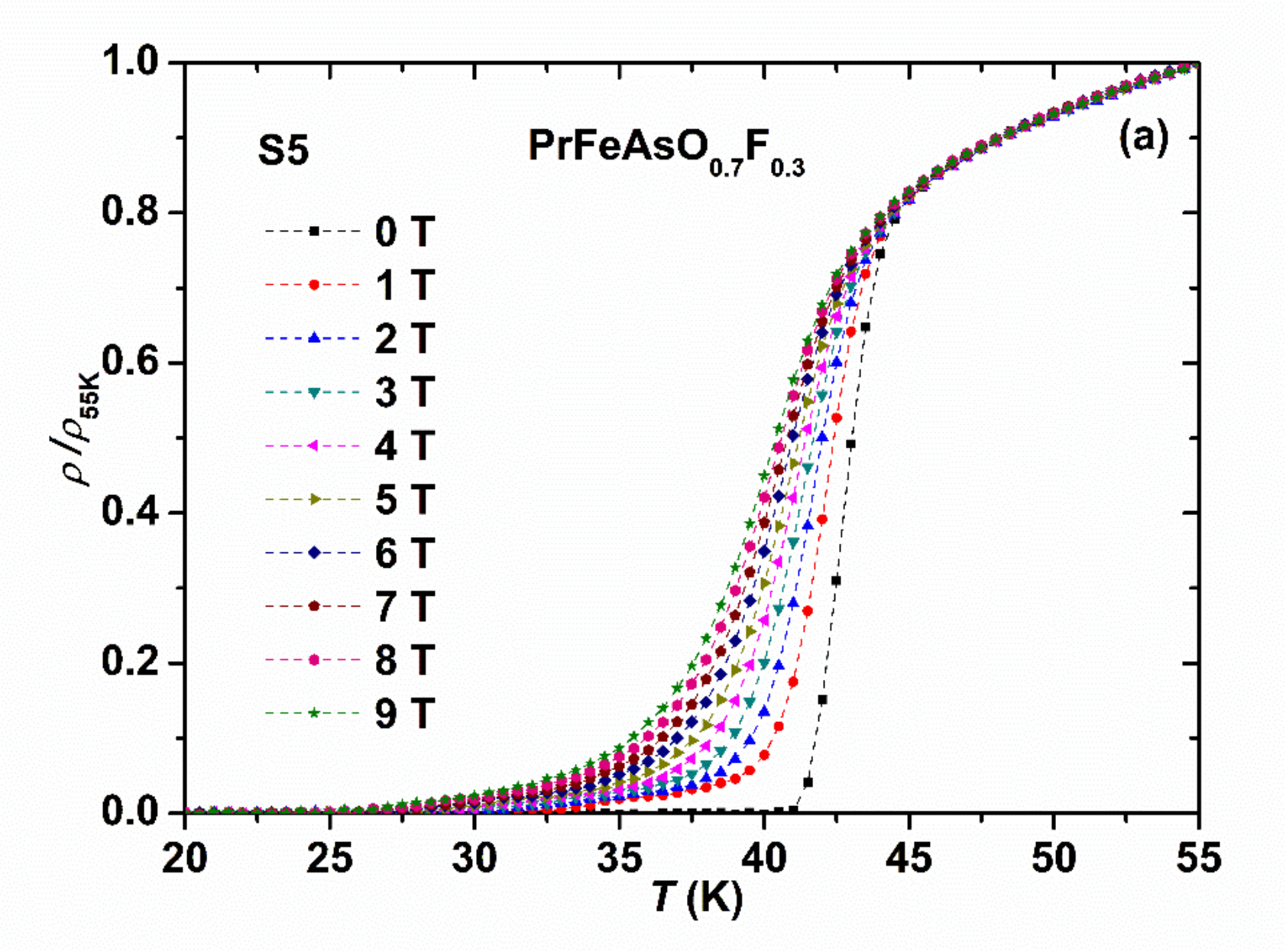

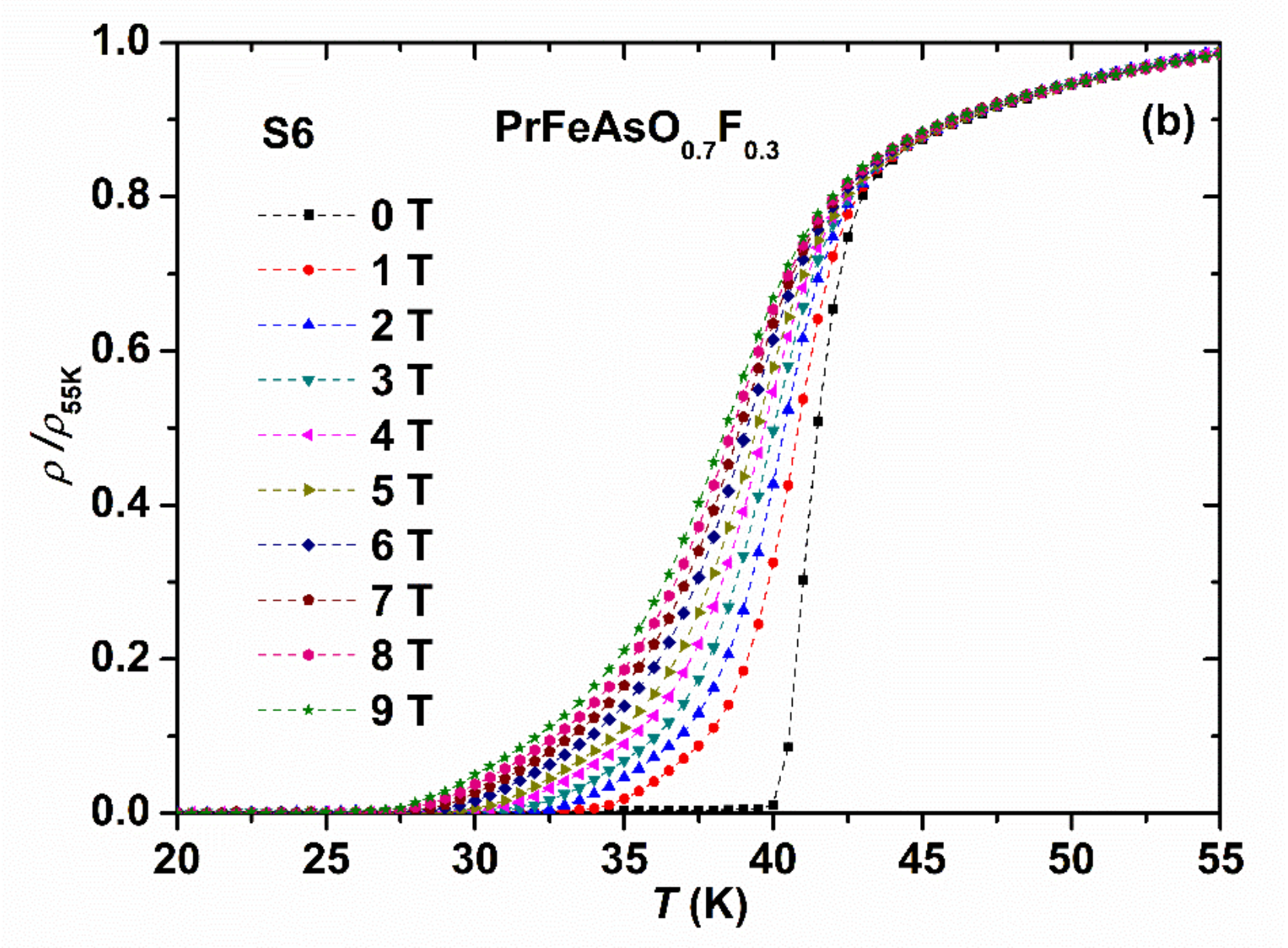

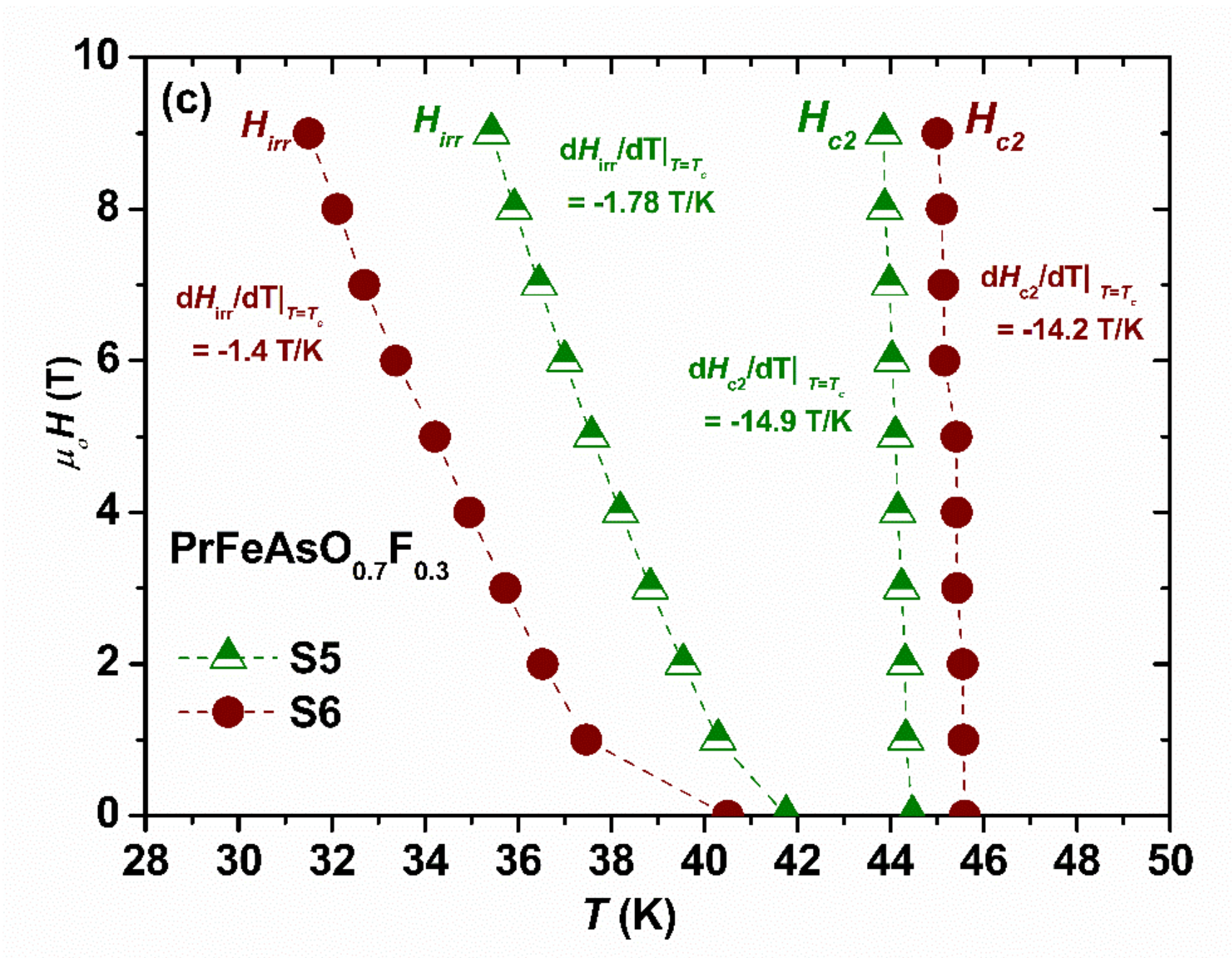
(c)
H_irr
H_irr
dH_irr/dT|T=Tc = -1.78 T/K
H_c2
H_c2
dH_irr/dT|T=Tc = -1.4 T/K
dH_c2/dT| T=Tc = -14.2 T/K
dH_c2/dT| T=Tc = -14.9 T/K
PrFeAsO0.7F0.3
S5
S6
μoH (T)
T (K)

**Figure 7:** **(a)** Arrhenius plots i.e. the variation of ln($\rho/\rho_{55K}$) versus $T^{-1}$ for the sample S5 under selected magnetic fields, with the dashed lines representing the linear fits in the thermally activated flux-flow (TAFF) regime. **(b)** Corresponding Arrhenius plots for the sample S6 under magnetic fields of 0–9 T. **(c)** Magnetic-field dependence of the vortex activation energy $U_0$ extracted from the TAFF analysis for the samples S5 and S6. The dashed lines represent power-law fits, $U_0 \propto H^{-\eta}$, with distinct low- (up to 7 T) and high-field (7 to 9 T) regimes characterized by the exponents $\eta_1$ and $\eta_2$.

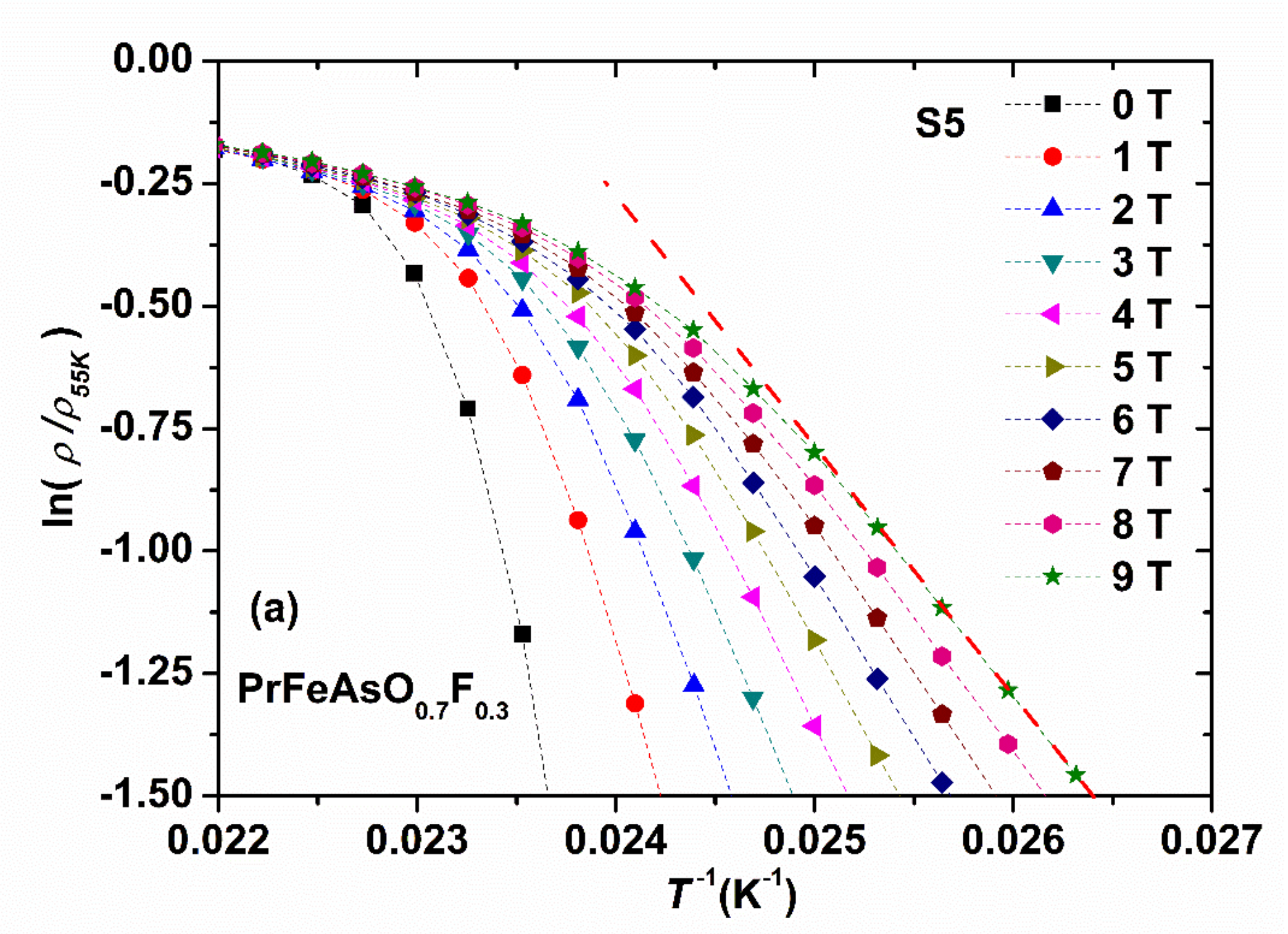

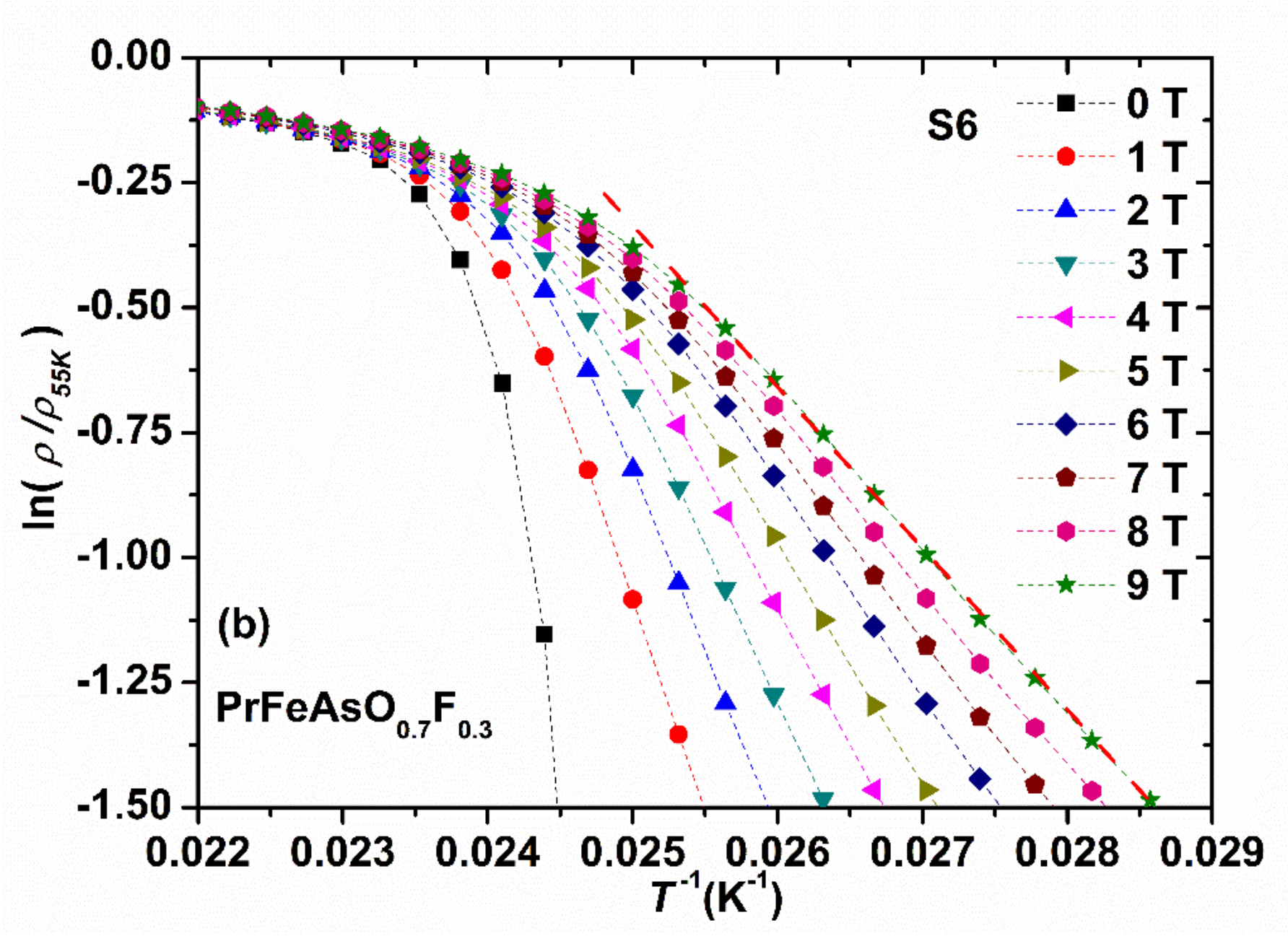

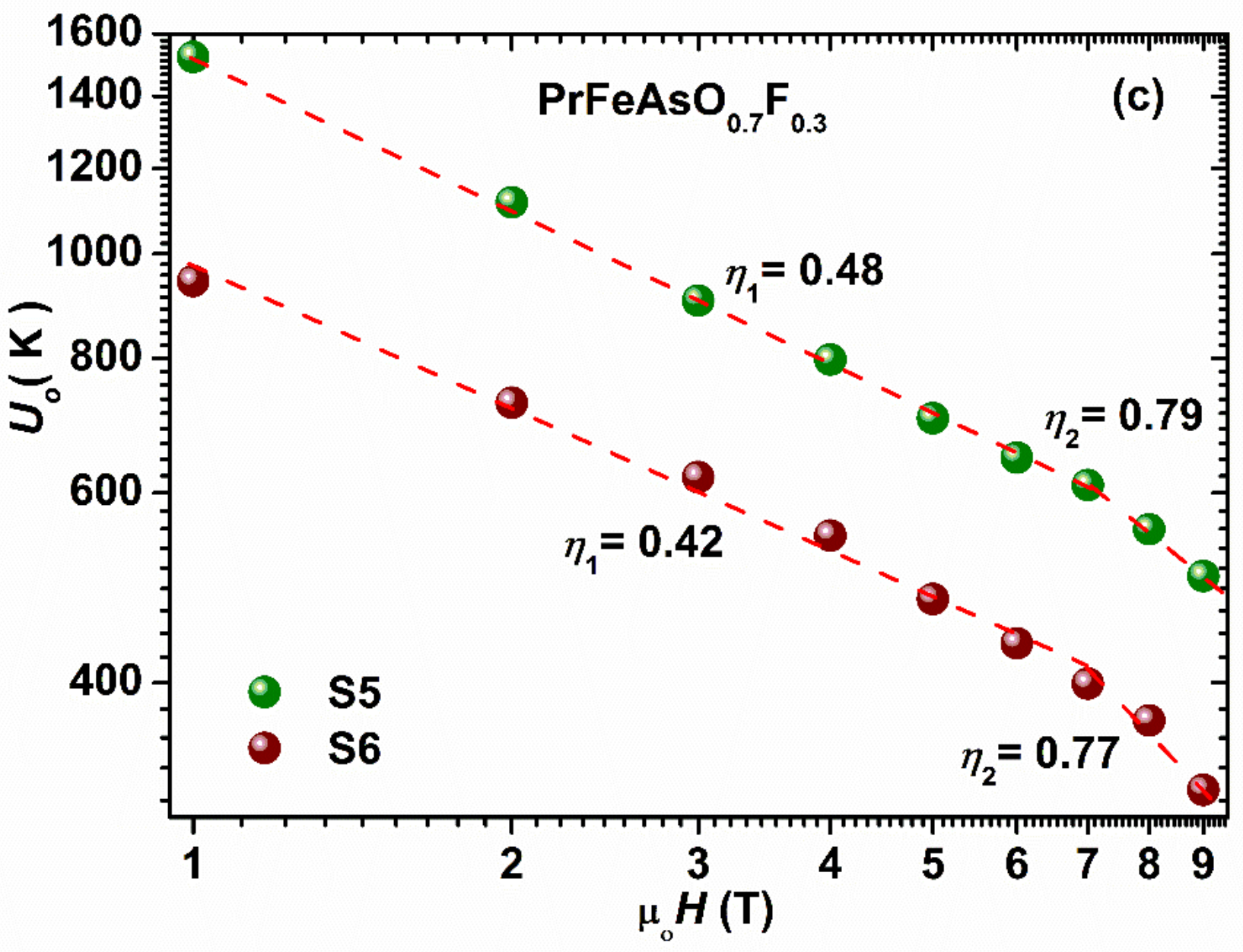

PrFeAsO0.7F0.3
(c)
η1= 0.48
η2= 0.79
η1= 0.42
η2= 0.77
S5
S6
Uo( K )
μoH (T)
1600
1400
1200
1000
800
600
400
1
2
3
4
5
6
7
8
9

**Figure 8:** (**a**) Temperature dependence of the normalized magnetization, $M/M_{5K}$, for the samples S1–S7 measured under zero-field-cooled (ZFC) and field-cooled (FC) conditions in an applied magnetic field of 20 Oe. (**b**) Magnetic-field dependence of the effective magnetic critical current density, $J_c$ calculated from the background corrected *M-H* hysteresis loops at 5 K using the Bean critical-state model [27] for the samples S1 to S7 at 5 K. The *M-H* loops for the samples S5 and S6 are shown in the Supplementary Figure Z12(a)-(b).

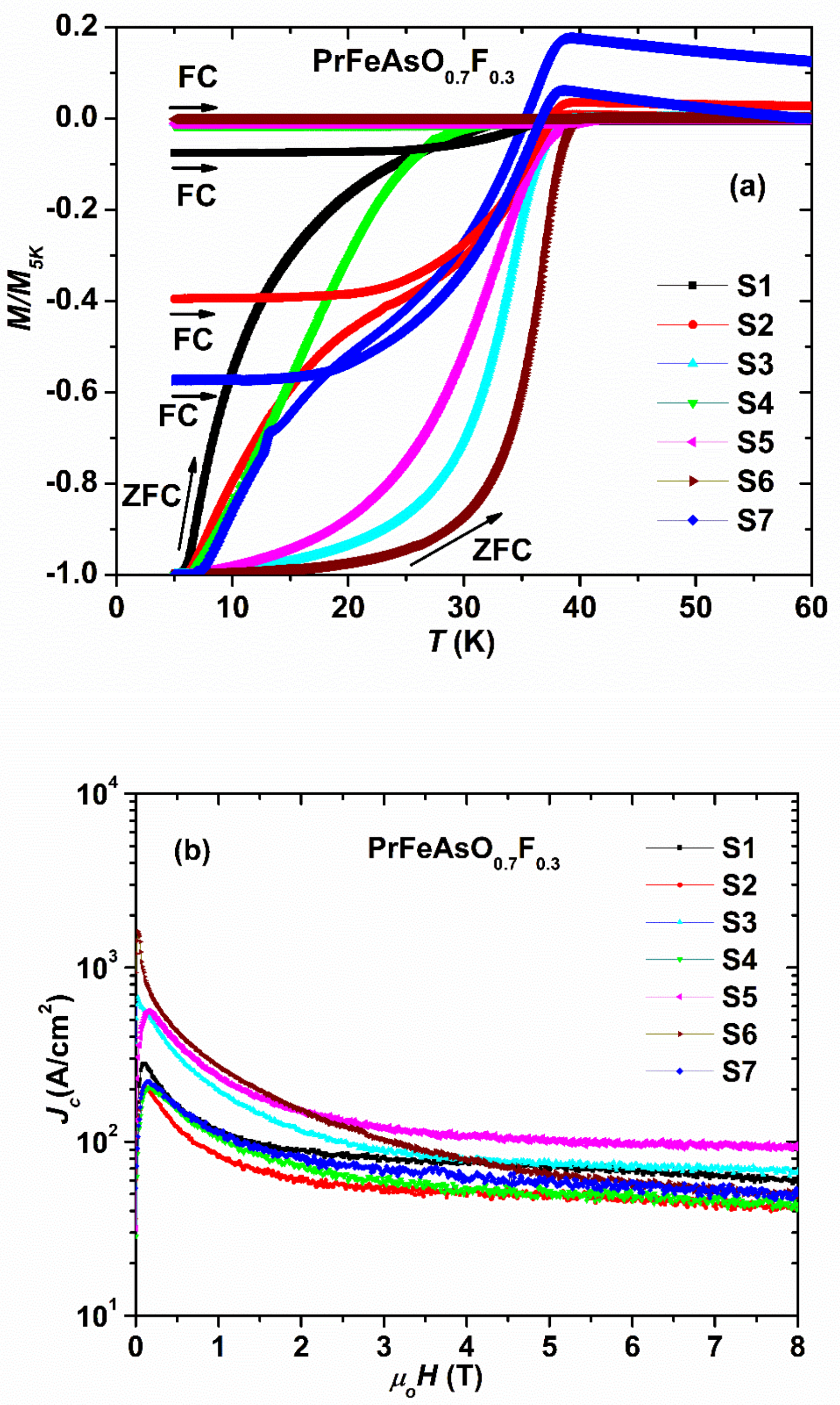

**Figure 9:** Pelletization-pressure dependence of the key superconducting and transport parameters of $PrFeAsO_{0.7}F_{0.3}$ samples: **(a)** the lattice parameter $a$, **(b)** the superconducting transition temperatures determined from the electrical resistivity ($T_c^{onset}$) and magnetization ($T_c^{mag}$) measurements, and **(c)** the magnetic critical current density $J_c$ measured at applied magnetic fields of 0.5 and 6 T at 5 K. Dashed curves are guides to the eye. The corresponding plots are provided in the Supplementary Figure Z11 without guide-to-the-eye lines and Figure Z13 with the EDX-derived actual fluorine content, $x_{act}$.

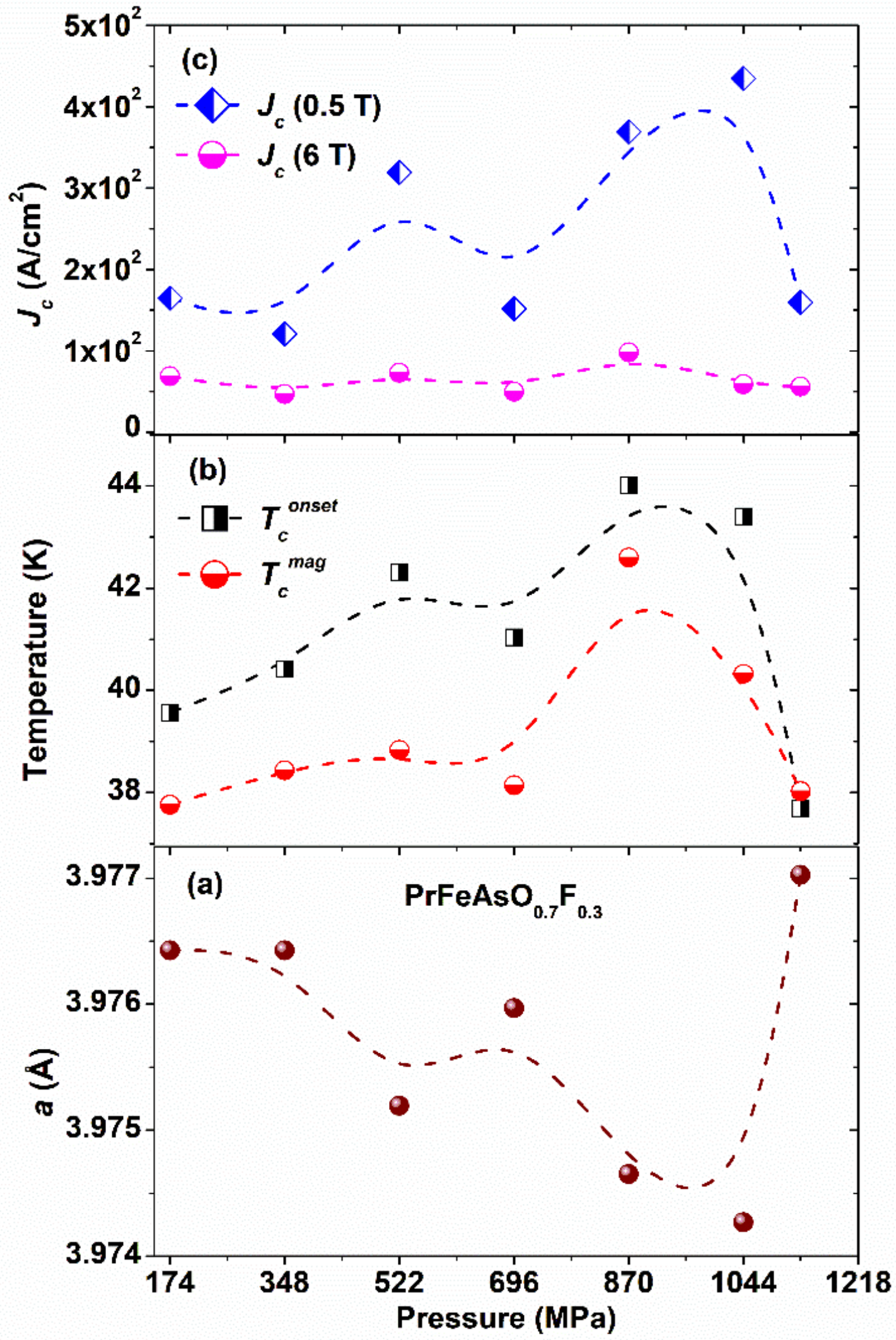